\documentclass[twocolumn,prb,multicol,amsmath,amssymb]{revtex4-2}
\usepackage[dvips]{graphicx}
\usepackage[bookmarks=false]{hyperref}
\usepackage{graphicx}
\usepackage{dcolumn}
\usepackage{bm}
\usepackage{graphics}
\usepackage{epsfig,color}
\usepackage[normalem]{ulem} 
\usepackage{soul}

\newcommand{\be}{\begin{equation}}
\newcommand{\ee}{\end{equation}}
\newcommand{\bea}{\begin{eqnarray}}
\newcommand{\eea}{\end{eqnarray}}

\begin{document}
\title {Exploring quantum coherence, spin squeezing and entanglement in an extended spin-1/2 XX Chain}
\author{S. Mahdavifar$^{1}$}
\email[]{smahdavifar@gmail.com}
\author{F. Khastehdel Fumani$^{2}$}
\author{B. Haghdoost$^{1}$}
\author{M. R. Soltani$^{3}$}
\affiliation{$^{1}$Department of Physics, University of Guilan, 41335-1914, Rasht, Iran}
\affiliation{$^{2}$Department of Basic Sciences, Langarud Branch, Islamic Azad University, Langarud, Iran}
\affiliation{ $^{3}$Department of Physics, Yadegar-e-Imam Khomeini (RAH), Shahr-e-Rey Branch, Islamic Azad University, Tehran, Iran}  

\begin{abstract}

In this study, we explore the ground state phase diagram of the spin-1/2 XX chain model, which features  $XZY-YZX$ type three-spin interactions (TSI). This model, while seemingly simple, reveals a rich tapestry of quantum behaviors. Our analysis relies on  several key metrics. The '$l_1$-norm of coherence' helps us identify coherent states within the phase diagram, which represent states capable of superposition and interference. We employ the 'spin squeezing parameter' to pinpoint unique coherent states characterized by isotropic noise in all directions, making them invaluable for quantum metrology. Additionally, we utilize the 'entanglement entropy' to determine which of these coherent states exhibit entanglement, indicating states that cannot be fully described by local variables. Our research unveils diverse regions within the phase diagram, each characterized by coherent, squeezed, or entangled states, offering insights into the quantum phenomena underling  these systems. We also study the critical scaling versus the system size for the mentioned quantities.

\end{abstract}
\maketitle

\section{Introduction}\label{sec1} 

Coherent states are quantum states that, in certain aspects, mimic classical states by possessing well-defined amplitudes and phases \cite{Coh1, Coh2, Coh3}. However, they also exhibit quintessentially quantum features like uncertainty and superposition. Quantum coherence, a fundamental concept in quantum physics, allows quantum systems to exist in superpositions of states, enabling interference. This concept has far-reaching applications in fields such as quantum optics, quantum information, quantum metrology, and quantum computing.

The framework for quantifying coherence in the context of quantum systems aims to measure a system's ability to exhibit superposition and interference, fundamental characteristics distinguishing quantum mechanics from classical physics. One approach to quantifying quantum coherence involves the use of interferometers, which are devices that split and recombine quantum waves, such as beams of light or electrons. By varying the path difference between the two branches of the interferometer, one can observe interference patterns on a detector. Another method employs correlation functions, mathematical tools for quantifying the relationships between two or more observables. For example, the quantum Fisher information, which is linked to a quantum state's sensitivity to small parameter changes  \cite{Coh4}, serves as a valuable measurement tool. From a theoretical standpoint, one proper measure of coherence is the l1-norm of coherence   \cite{Coh5, Coh6}:

\begin{eqnarray}\label{eq-l1}
C_{l_1}(\hat{\rho})=\sum_{i\neq j} \big |\rho_{ij} \big |,
\end{eqnarray}
 where  $\hat{\rho}$ is the density operator of the quantum state, and $\rho_{ij}=\big< \psi_i| \hat{\rho} |\psi_j \big>$. $\{|\psi_m \big> \}$ are basis kets of the Hilbert space. It is important to note that density matrices that are diagonal in this basis are considered incoherent. The $l1$-norm of coherence depends on the choice of basis, as different bases may have different off-diagonal elements. Therefore, the $l1$-norm of coherence is not an invariant quantity under basis transformations.

Coherent states are also defined based on the spin-squeezing parameter (SSP)   \cite{Sq1,Sq2,Sq3,Sq3-0, Sq3-1, Sq4}. The SSP quantifies the extent to which a quantum state, composed of an ensemble of spins, deviates from a coherent state. In the context of spin-squeezed states, a unique type of quantum coherent state, there's a reduction in uncertainty regarding one angular momentum component. However, this comes at the cost of increased uncertainty in another component. From the perspective of SSP, coherent states exhibit uniform noise in all directions, rendering them more responsive to rotations around a specific axis. This property makes SSP a valuable tool for high-precision metrology, a field dedicated to enhancing the precision and accuracy of measurements using quantum phenomena  \cite{I5,I6,I7,I8,I9,ASq1}.

Spin squeezing finds numerous applications in quantum physics. For example, it proves useful in quantum metrology, where quantum effects are harnessed to improve measurement precision. It can also be applied to the study of quantum phase transitions  \cite{12n, 13n, 14n, 15n, 16n, 17n,18n,18n0, 18n1, 19n, 20n}, which entail sudden changes in a system's properties due to quantum fluctuations.

\begin{figure}
\centerline{\includegraphics[width=0.5\linewidth]{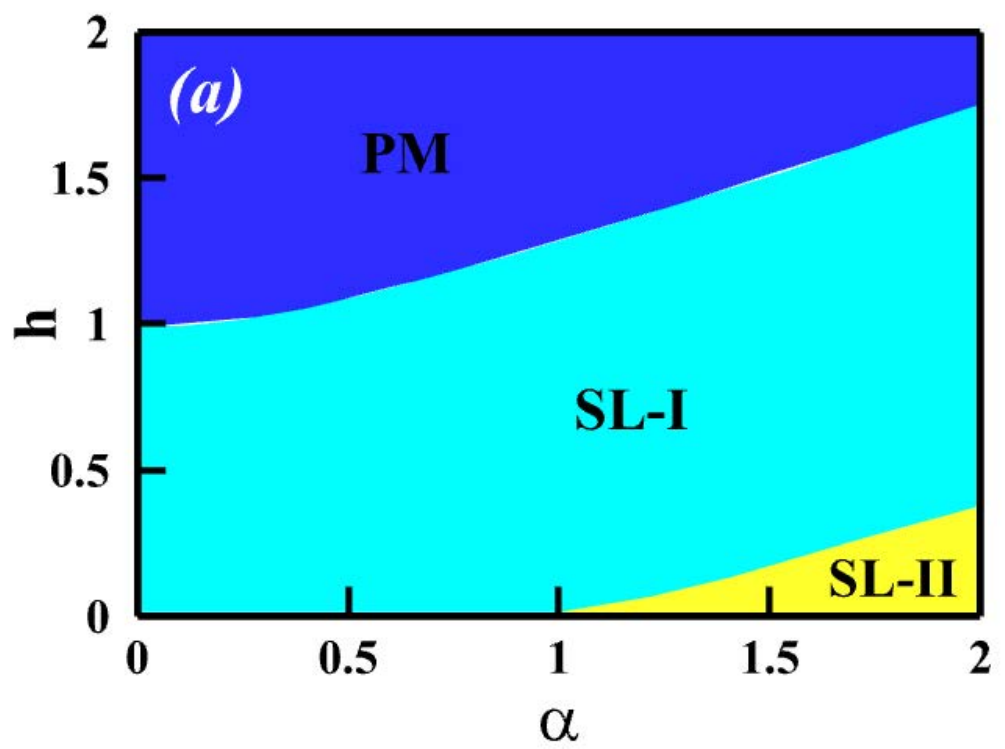}  \includegraphics[width=0.55\linewidth]{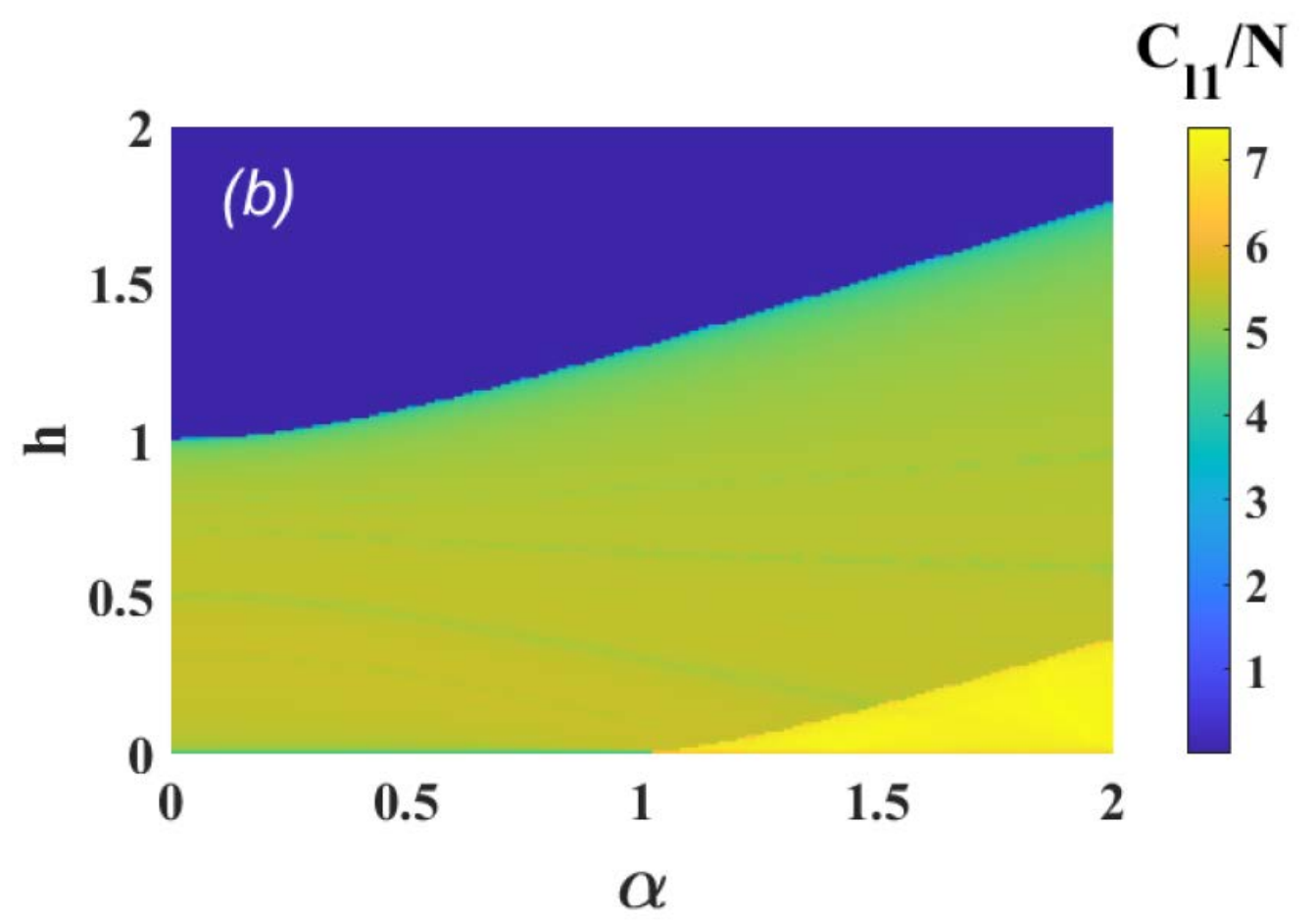}  }
\centerline{\includegraphics[width=0.55\linewidth]{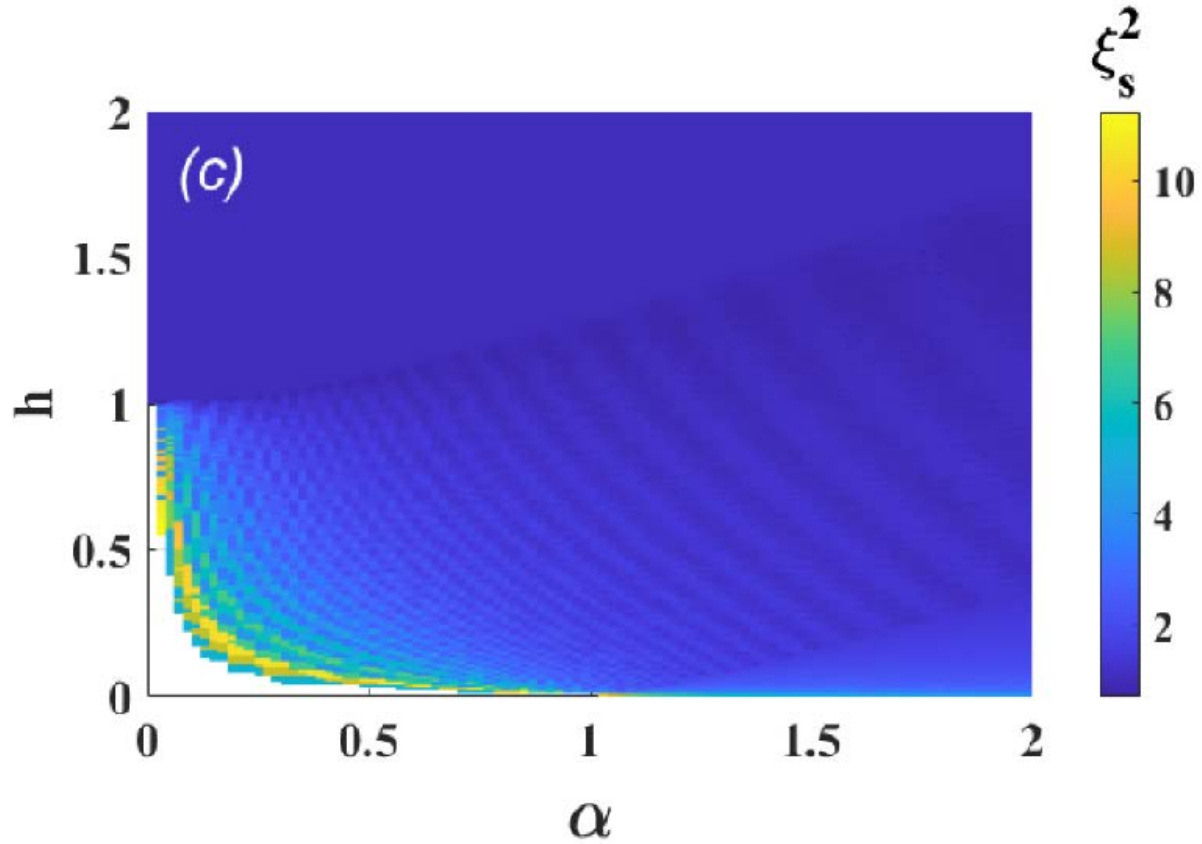} \includegraphics[width=0.55\linewidth]{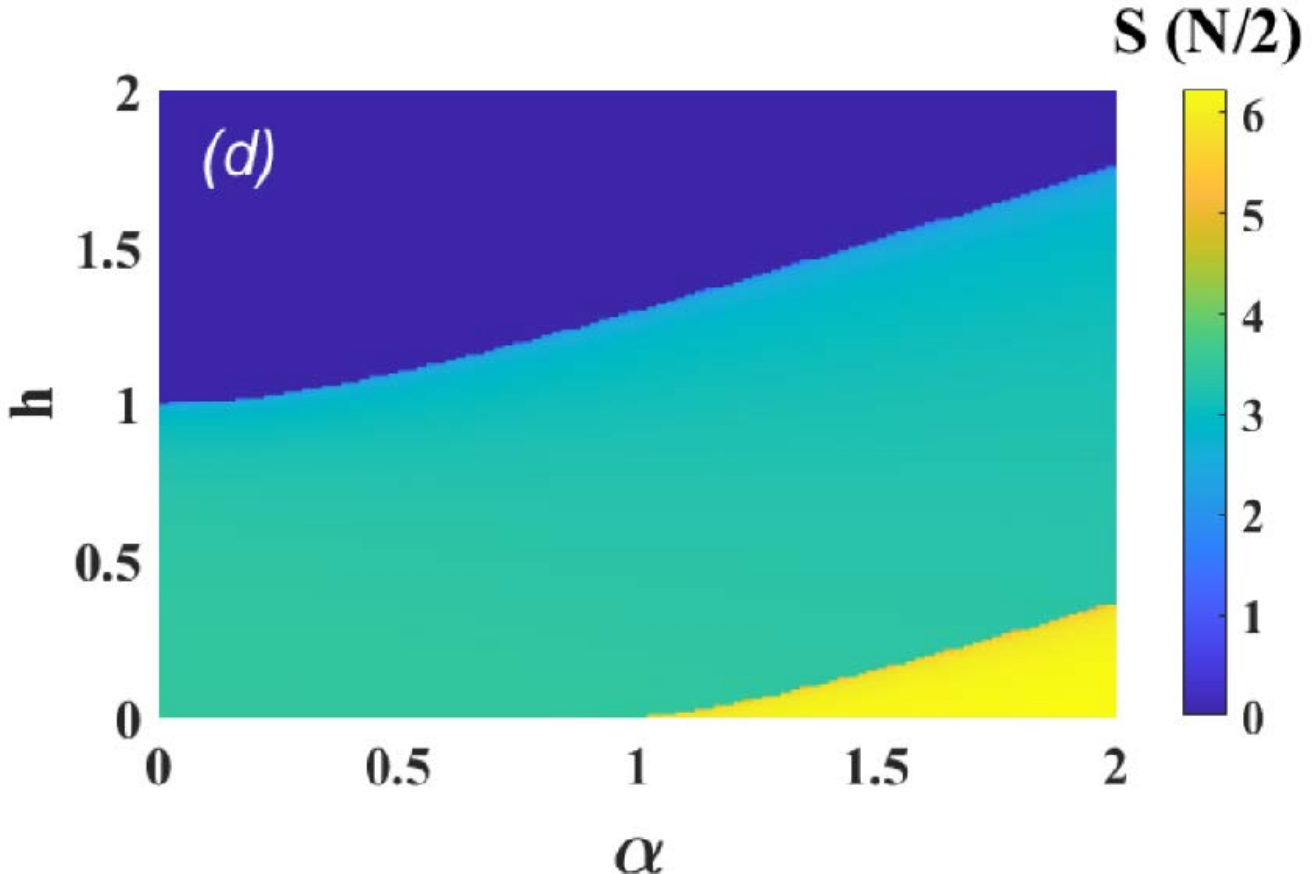}  }
\caption{(color online).  (a) The ground state phase diagram of the model. The density plot of (a) the $l1$-norm of coherence, (c) the SSP and (d) the EE parameter for a chain size $N=500$.}
\label{Fig0}
\end{figure}

SSP can be defined in various ways depending on the context and the intended application. A commonly used definition is based on the Heisenberg uncertainty relation for angular momentum components, which sets a lower limit on the product of their variances.  Two measures of SSP are the Kitagawa-Ueda  parameter \cite{Sq2} and the Wineland parameter \cite{Sq3}. The Kitagawa-Ueda parameter is a suitable measure of spin squeezing for spin systems with a well-defined mean spin direction and a large number of particles. In contrast, the Wineland parameter is suitable for spin systems that are sensitive to $SU(2)$ rotations and have a small number of particles \cite{Sq3-0, Sq3-1}.   The Kitagawa-Ueda SSP  is introduced as   \cite{Sq2}:

\begin{equation}\label{squeez}
\xi _s^2 = \frac{{4{{({\Delta {J_{\vec{n}_{\perp}}}})}^2}}}{N}.
\end{equation}
Here, $\xi _s^2$ is the spin squeezing parameter, $\vec{n}_{\perp}$ represents the axis perpendicular to the average spin direction $\vec{n}_0$, and  the variance $(\Delta J)^2$ is minimized. The total spin components $J_{\alpha}$ of $N$  particles satisfy the commutation relation  $[J_{\alpha},J_{\beta}]=i \hbar J_{\gamma}$, where $\alpha, \beta, \gamma$ are cyclic permutations of $x, y, z$. Another SSP  is $\xi_R^2 = \frac{{N{{({\Delta {J_{\vec{n}{\perp}}}})}^2}}}{      \big | \big< J_{\vec{n}} \big> \big|   }$, which was introduced by Wineland  \cite{Sq3}. In this context, a coherent state is characterized by $\xi_s^2 =1$, whereas a spin-squeezed state exhibits  $\xi _s^2 <1$.

Both definitions of the SSP are closely linked to the concept of quantum entanglement   \cite{co0, co1, co2, co3, co4, co5, co6}. Quantum entanglement represents a phenomenon in which two or more quantum systems share correlations that cannot be explained by classical physics. Spin-squeezed states are illustrative examples of entangled states because they exhibit correlations among different spins that exceed the capabilities of classical states.
It is important to note that entanglement leads to the emergence of entropy, specifically "entanglement entropy" (EE)  \cite{27n0}. EE serves as a measure of the extent of quantum entanglement between two subsystems of a composite quantum system. It quantifies the information loss or inaccessibility when one only has access to one of the subsystems and not the entire system. One common approach to defining EE is based on the concept of the density operator, and it involves calculating the von Neumann entropy of a reduced density matrix for a subsystem  \cite{28n,29n,30n}. 
For a bipartite system, the EE in the pure ground state $\mid \psi >$, represented by the density matrix $\rho=$ $\mid \psi >$ $ < \psi \mid$, is defined as the von Neumann entropy of subsystem $A$:
\begin{equation}
S_{A}=-\bf {Tr}[ \rho_{A} \log_{2} (\rho_{A})],
 \end{equation}

where $\rho_{A}$ is the reduced density matrix of $A$, obtained by tracing over the rest of the system, $B$:
\begin{equation}
\rho_{A}= \bf {Tr}_{B} (\rho).
 \end{equation}

EE typically scales with the boundary area of subsystem $A$, rather than its volume. This behavior is distinct from what one might expect and is known as the 'area law' for EE. It has been the subject of extensive study in recent years. In noncritical ground states of spin chains with a finite correlation length, the EE remains constant. However, at a quantum critical point, when subsystem $A$ is a finite interval of length $l$, the EE slightly deviates from the area law due to a logarithmic correction:
\begin{equation}
S_{A} (l) \sim \frac{c}{3} \log(l),
 \end{equation}
where  $c$ represents the central charge  \cite{31n,32n}.

In this paper, we delve into the study of the exactly solvable spin-1/2 XX chain model with a Three-Spin Interaction (TSI), examining its behavior in the presence of a magnetic field   \cite{Ne-1, Ne-2}. The system exhibits intriguing characteristics, including two distinct gapless spin liquid phases known as SL-I and SL-II, in addition to the gapped paramagnetic (PM) phase.
Our approach begins with the application of the Fermionization technique to derive the ground state of the system. Subsequently, we utilize the $l_1$-norm function as a robust framework for quantifying coherence. Using this method, we calculate the SSP and the EE across the entire ground state phase diagram of the model.  We have obtained the critical lines for the system by using these functions (See Fig.~\ref{Fig0}). We have found a squeezed region in the presence of a magnetic field, where the system is in the SL-II phase. We have also detected some coherent states by examining the SSP and verifying them with the $l_1$-norm of coherence. Moreover, we have shown that the EE exhibits critical scaling in both SL-I and SL-II phases.

The paper is structured as follows:
In the next section (Section II), we introduce the model and the analytical tools used in our study, providing a foundation for understanding our methodology.
In Section III, we present our research findings related to the Spin Squeezing (SS), the $l_1$-norm of coherence, and the EE, offering a detailed analysis of our results.
Finally, in Section IV, we provide a conclusion and a comprehensive summary of our key findings and their implications.


We study the spin-1/2 XX chain model with a three-spin interaction (TSI) in a magnetic field, which is described by the Hamiltonian

\begin{eqnarray}\label{eq1}
{\cal H} &=&-J\sum\limits_{n = 1}^N {\left( {S _n^x S _{n + 1}^x + S _n^y S _{n + 1}^y} \right)} \nonumber \\
&-&J^*\sum\limits_{n = 1}^N S _{n +1}^z{\left( {S _n^x S _{n + 2}^y - S _n^y S _{n +2}^x} \right)} \nonumber \\
&-& J h\sum\limits_{n = 1}^N {S _n^z}~,
\end{eqnarray}
where $S_n$ is the spin operator at site $n$, $J>0$ is the ferromagnetic (FM) exchange coupling, $J^*$ and $h$ are the TSI strength and the magnetic field intensity, respectively, and $N$ is the number of spins. We assume a periodic boundary condition, such that $S_{N+1}^\mu=S_1^\mu$ ($\mu=x,y,z$). We also define $\alpha=\frac{J^*}{J}$ for convenience.

The spin-1/2 XX chain model with the TSI interaction has a rich ground state phase diagram, which depends on the ratio $\alpha=\frac{J^*}{J}$ of the TSI strength and the FM exchange coupling. When the magnetic field is zero, the phase diagram is known exactly \cite{Ne-2}. It contains two gapless spin liquid (SL) phases, SL-I and SL-II, which are separated by a second-order quantum phase transition at $\alpha_c=1$.
This quantum phase transition is notable for the doubling of Fermi points in the representation of spinless fermions, providing a convenient means of mapping the spin model to a free fermion model. When a magnetic field is introduced, the system undergoes a second-order phase transition from the SL-I phase to the paramagnetic (PM) phase. If the system is initially in the SL-II phase, it first transitions to the SL-I phase at the first critical field and then to the PM phase at a higher second critical field. Importantly, both transitions are of second order.
 This model has garnered significant attention in recent years, leading to numerous studies exploring its properties and applications \cite{Ne-3,Ne-4,Ne-5,Ne-6,Ne-7}.
For instance, a connection was established between the cooling-heating efficiency and the ground state phase diagram of the model. It was revealed that the efficiency could be enhanced by tuning the strength of the TSI and the magnetic field \cite{Ne-3}.

In a separate study \cite{Ne-4}, the authors examined the ground state phase diagram of the model using the Lee-Yang zeros, order parameters, and ground-state energy. Their research demonstrated the presence of chiral ordering within the SL-II phase. Furthermore, the location of quantum critical lines was determined using quantum discord \cite{Ne-5}.
Additionally, it was shown that steered quantum coherence and its first-order derivative effectively serve as indicators for different critical points within the ground state phase diagram  \cite{Ne-6}.

In $2009$, it was suggested that a system of trapped ions could be a promising avenue to explore the unique characteristics of spin models with three-body interactions \cite{Ne-8}. This idea has become a reality in practice, as by $2023$, experiments have demonstrated a new class of native interactions between trapped-ion qubits, extending beyond conventional pairwise interactions to include three- and four-body spin interactions  \cite{Ne-9}.


In another line of research, the Mølmer-Sørensen scheme has been extended to induce three-spin interactions through tailored first- and second-order spin-motion couplings \cite{Ne-10}. This extended scheme allows for the engineering of single-, two-, and three-spin interactions and can be finely tuned using an enhanced protocol to faithfully simulate purely three-spin dynamics.
Additionally, a significant experimental effort in Ref.~ \cite{Ne-11} involved simulating a particularly intriguing system with competing one-, two-, and three-body interactions. The study observed the emergence of distinct ground states within this complex system.

Here we continue the study on the ground state of this model. To diagonalize the Hamiltonian, we first use the Jordan-Wigner transformation, which maps the spin operators to fermionic operators as follows:

\begin{eqnarray}
S^{+}_{n}&=&c_{n}^{\dag}e^{i\pi\sum^{n-1}_{m=1}c^{\dag}_{m}c_{m}},\nonumber\\
S^{-}_{n}&=&e^{-i\pi\sum^{n-1}_{m=1}c^{\dag}_{m}c_{m}}c_{n},\nonumber\\
S^{z}_{n}&=&c_{n}^{\dag}c_{n}-\frac{1}{2},
\label{eq17}
\end{eqnarray}
where $c_n^\dagger$ and $c_n$ are the fermionic creation and annihilation operators. This transforms the Hamiltonian to a quadratic form in terms of fermionic operators:
\begin{eqnarray}\label{eq2}
{\cal H} &=&-\frac{J}{2} \sum\limits_{n = 1}^N \left(c_n^\dag c_{n + 1}+c_{n+1}^\dag c_n      \right) \nonumber\\
&+&i \frac{J^*}{4} \sum\limits_{n = 1}^N \left(c_n^\dag c_{n + 2}-c_{n+2}^\dag c_n      \right) \nonumber\\
&-&J h \sum\limits_{n = 1}^N c_n^\dag c_{n }.
\end{eqnarray}
Then, we use the Fourier transformation ${c_n} =  \sum_k e^{ - ikn} {c_k}$, which diagonalizes the Hamiltonian in terms of fermionic modes with wave number $k$: 
\begin{equation}\label{eq3}
{\cal H} = \sum\limits_k \varepsilon_k~ c_k^\dag c_{k },
\end{equation}
with energy spectrum 
\begin{equation} \label{spectrum}
\varepsilon_k = -J(h+\cos(k)-\frac{\alpha}{2} \sin(2 k)).
\end{equation}
It should be noted that the summation in Eq. (\ref{eq3}) runs over $k=2\pi m/N$, with \\
$m=0,\pm 1,...,\pm \frac{1}{2}(N-1)[m= 0, \pm 1,..., \pm (\frac{1}{2}N-1), \frac{1}{2}N]$ for $N$ odd [$N$ even] (having imposed periodic boundary conditions on the JW fermions). In the thermodynamic limit $N\longrightarrow \infty$, the ground state of the system corresponds to the configuration
where all the states with $\varepsilon_k<0$ are filled and $\varepsilon_k>0$ are empty and written as 
\begin{equation} \label{spectrum}
\big | GS \big >=\prod_{\Lambda} c_k^\dag \big | 0 \big >,
\end{equation}
where $\Lambda$ is a region in the $k$-space where $\varepsilon_k<0$ and $\big | 0 \big >$ denotes the vacuum state.

The correlation matrix is a valuable tool for investigating fermionic systems because it reflects the single-particle features and correlations of the many-body wave function \cite{33n,33n-0,33n-1,33n-2,34n}. The density matrix of the ground state is a matrix that describes the probability distribution of system configurations when the system is in its minimum energy state. Both the density matrix and the correlation matrix are derived from the same ground state and share the same eigenvalues. However, it's important to note that this doesn't imply the density matrix and the correlation matrix have identical off-diagonal elements. As mentioned earlier, the $l_1$-norm of coherence is the sum of absolute values of the off-diagonal elements of the density matrix. A non-zero value of the $l_1$-norm indicates that the ground state is coherent. It's crucial to understand that only the non-zero property is relevant, not the exact value. For this reason, we approximate the $l_1$-norm of coherence in the basis of the $z$-component of the total spin with the sum of absolute values of off-diagonal elements of the correlation matrix. In the following, we demonstrate that this approximation reveals very important features of the system.

The correlation matrix of the ground state is a matrix of expectation values of fermionic operators,
\begin{equation}\label{eq6}
\rho_N= \left( {\begin{array}{*{20}{c}}
\langle c_{1}^{\dag} c_{1}\rangle& \cdots&\langle c_{1}^{\dag} c_{N}\rangle&\langle c_{1}^{\dag} c_{1}^{\dag}\rangle&\cdots&\langle c_{1}^{\dag} c_{N}^{\dag}\rangle\\
\langle c_{2}^{\dag} c_{1}\rangle& \cdots&\langle c_{2}^{\dag} c_{N}\rangle&\langle c_{2}^{\dag} c_{1}^{\dag}\rangle&\cdots&\langle c_{2}^{\dag} c_{N}^{\dag}\rangle\\
\vdots&\vdots&\vdots&\vdots&\vdots&\vdots&\\
\langle c_{l}^{\dag} c_{1}\rangle&\cdots&\langle c_{l}^{\dag} c_{N}\rangle&\langle c_{l}^{\dag} c_{1}^{\dag}\rangle&\cdots&\langle c_{l}^{\dag} c_{N}^{\dag}\rangle\\
\langle c_{1} c_{1}\rangle&\cdots&\langle c_{1} c_{N}\rangle&\langle c_{1} c_{1}^{\dag}\rangle&\cdots&\langle c_{1} c_{N}^{\dag}\rangle\\
\langle c_{2} c_{1}\rangle&\cdots&\langle c_{2} c_{N}\rangle&\langle c_{2} c_{1}^{\dag}\rangle&\cdots&\langle c_{2} c_{lN}^{\dag}\rangle\\
\vdots&\vdots&\vdots&\vdots&\vdots&\vdots\\
\langle c_{l} c_{1}\rangle&\cdots&\langle c_{l} c_{N}\rangle&\langle c_{l} c_{1}^{\dag}\rangle&\cdots&\langle c_{l} c_{N}^{\dag}\rangle
\end{array}} \right),\\
\end{equation}
where
\begin{eqnarray}\label{eq55}
\langle c_{n}^{\dag} c_{m} \rangle&=&\frac{1}{N} \sum\limits_{k \in \lambda} e^{-i(k(m-n))},\nonumber\\
\langle c_{n}^{\dag} c_{m}^{\dag} \rangle&=&0.
\end{eqnarray}

From this correlation matrix, the $l_1$-norm of coherence is approximated as

\begin{eqnarray}\label{eq-l1}
C_{l_1}(\hat{\rho}) \simeq \sum_{m \neq n}\big |  \langle c_{n}^{\dag} c_{m}\rangle \big |.
\end{eqnarray}

On the other hand, the EE of a finite block of $l$ sites in an infinite system of free spinless fermions can be computed by \cite{33n,34n}
\begin{eqnarray}
S_{A}=-\sum\limits_{\gamma = 1}^{2l} C_{\gamma} \log(C_{\gamma}),
\end{eqnarray} 
where $C_{\gamma}$ is one of the $2 l$ eigenvalues of the density matrix $\rho_{l}$. Considering the symmetries of the model as the unbroken $Z_2$ invariance for finite $N$ implies that 
\begin{eqnarray}\label{eq6}
\left\langle {J_x } \right\rangle  = \left\langle {J_y } \right\rangle  = 0,
\end{eqnarray}
and similarly,
\begin{equation}
\left\langle {J_\alpha J_z} \right\rangle  = \left\langle {J_z J_\alpha } \right\rangle  = 0, \ \ \alpha  = x,y.
\end{equation}
The magnetization for $h>0$ is always along the $z$-axis, with full polarization developing in the PM phase. As a result, 
$J_{\vec{n}_{\perp}} = \cos (\Omega )J_x + \sin (\Omega)J_y$, with $\Omega$ to be chosen  to minimize
\begin{eqnarray}\label{eq7}
({\Delta {J_{\vec{n}_{\perp}}}})^2 &=& \left\langle (J_ {\overrightarrow n_\bot } )^2 \right\rangle  - \left\langle J_ {\overrightarrow n_\bot }  \right\rangle ^2 \nonumber \\
&=& \langle (\cos(\Omega)J_x + \sin(\Omega)J_y)^2\rangle.
\end{eqnarray}
One can easily show that
\begin{eqnarray} \label{eq8}
\xi _s^2 &=& \frac{2}{N}\mathop {\min }\limits_\Omega  \big( {\left\langle {J_x^2 + J_y^2} \right\rangle } + \cos (2\Omega )\left\langle {J_x^2 - J_y^2} \right\rangle    \nonumber \\
&+& \sin (2\Omega) \left\langle {J_x J_y + J_y J_x} \right\rangle  \big),  \nonumber \\
&=&\frac{2}{N} \big [\left\langle {J_x^2 + J_y^2} \right\rangle - \sqrt{      \left\langle {J_x^2 - J_y^2} \right\rangle^{2}+ \left\langle {J_x J_y + J_y J_x} \right\rangle^{2}       }\big ]. \nonumber \\
\end{eqnarray}   
Finally, using the definition of the total spin of the particles, the SS parameter is obtained as
\begin{eqnarray} \label{eq8}
\xi _s^2 &=& 1+2 \sum\limits_{n = 1}^{N-1}(G_{n}^{xx}+G_{n}^{yy})\nonumber \\
&-&2\sqrt{ \big[\sum\limits_{n = 1}^{N-1}(G_{n}^{xx}-G_{n}^{yy})\big]^{2} + \big[\sum\limits_{n = 1}^{N-1}(G_{n}^{xy}+G_{n}^{yx})\big]^{2}           }, \nonumber \\
\end{eqnarray}   
where $G_{n}^{\alpha \beta}$ denotes two-point correlation function. Introducing $A_n = a_n^\dagger + a_n$ and $B_n = a_n^\dagger - a_n$, a direct calculation shows that
\begin{eqnarray}  \label{t1}
G_n^{xx } &\!=\!& \langle S_1^xS_{1 + n}^x \rangle  = \frac{1}{4}\langle B_1 A_2B_2...A_nB_nA_{n+1} \rangle ,   \nonumber \\
G_n^{yy} &\!=\!&\langle S_1^yS_{1+ n}^y \rangle  = \frac{( - 1)^n}{4}\langle A_1B_2A_2...B_nA_nB_{n+1} \rangle , \nonumber \\ 
G_n^{xy } &\!=\!&\langle S_1^xS_{1 + n}^y \rangle  = \frac{ - i}{4}\langle B_1A_2B_2...A_nB_nB_{n+1} \rangle \nonumber , \\
G_n^{yx } &\!=\!&\langle S_1^yS_{1 + n}^x \rangle  = \frac{  i( - 1)^n}{4}\langle A_1B_2A_2...B_nA_nA_{n+1} \rangle. \nonumber \\ 
\end{eqnarray}
These equations may be written in the generic form,
\begin{eqnarray} \label{eq2n}
G_n^{\alpha \beta}  = D_n^{\alpha \beta }\left\langle {{\phi _1}{\phi _2}{\phi _3}...\phi _{2n-2}\phi _{2n-1}{\phi _{2n}}} \right\rangle,
\end{eqnarray}
with
\begin{eqnarray}
D_n^{xx} &\!=\!& \frac{1}{4}, \ \ \ D_n^{yy} = \frac{( - 1)^n}{4}, \nonumber \\
D_n^{xy} &\!=\!& \frac{ - i}{4}, \ \ \ D_n^{yx} \!=\! \frac{i( - 1)^n}{4},
\end{eqnarray}
where each operator $\phi_j, j= 1,2, \ldots, 2n$, is identified with either an $A_n$ or a $B_n$ operator. Using the Wick theorem \cite{27n}, the $2n$-point functions can be expressed as Pfaffians  
\begin{multline} \label{pf2n}
G_n^{\alpha \beta} 
=  D_n^{\alpha \beta } \mbox{pf} \left( {\begin{array}{*{20}{c}}
{\left\langle {{\phi _1}{\phi _2}} \right\rangle }&{\left\langle {{\phi _1}{\phi _3}} \right\rangle }&{\left\langle {{\phi _1}{\phi _4}} \right\rangle }& \cdots &{\left\langle {{\phi _1}{\phi _{2n}}} \right\rangle }\\
{}&{\left\langle {{\phi _2}{\phi _3}} \right\rangle }&{\left\langle {{\phi _2}{\phi _4}} \right\rangle }& \cdots &{\left\langle {{\phi _2}{\phi _{2n}}} \right\rangle }\\
{}&{}&{\left\langle {{\phi _3}{\phi _4}} \right\rangle }& \cdots &{\left\langle {{\phi _3}{\phi _{2n}}} \right\rangle }\\
{}&{}&{}& \ddots & \vdots \\
{}&{}&{}&{}&{\left\langle {{\phi _{2n - 1}}{\phi _{2n}}} \right\rangle }
\end{array}} \right)\\
\end{multline}
where we have written the skew-symmetric matrix in the standard abbreviated form. One can easily find,

\begin{eqnarray}\label{eq56}
\langle A_{n} A_{m} \rangle&=& \delta(m-n)- \frac{2i}{N}  \sum\limits_{k \in \lambda} \sin(k(m-n))   ,\\
\langle B_{n} B_{m} \rangle&=& -\delta(m-n)+ \frac{2i}{N}  \sum\limits_{k \in \lambda} \sin(k(m-n))   ,\nonumber\\
\langle A_{n} B_{m} \rangle&=& \delta(m-n) - \frac{2}{N}  \sum\limits_{k \in \lambda} \cos(k(m-n))  ,\nonumber\\
\langle B_{n} A_{m} \rangle&=& -\delta(m-n) + \frac{2}{N}  \sum\limits_{k \in \lambda} \cos(k(m-n)).\nonumber
\end{eqnarray}

\begin{figure}
\centerline{\includegraphics[width=0.55\linewidth]{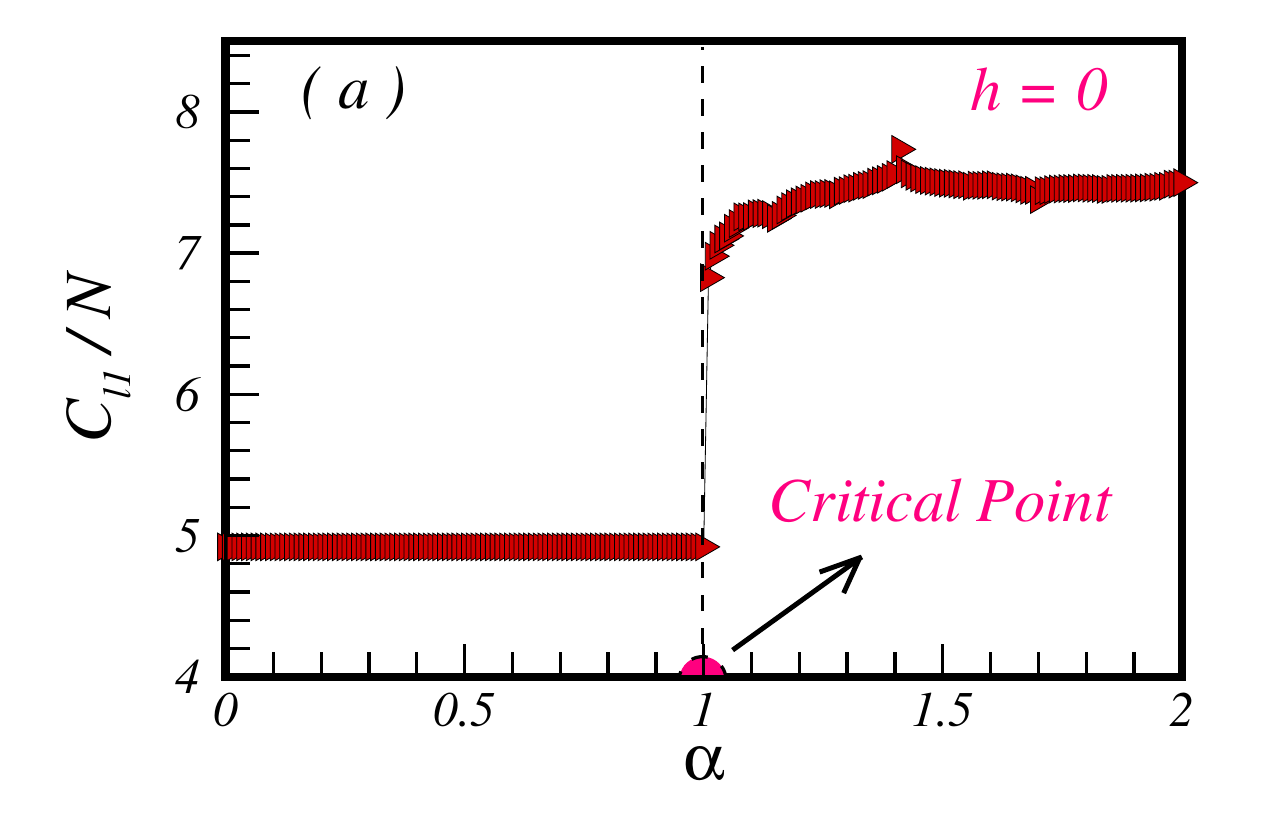}  \includegraphics[width=0.55\linewidth]{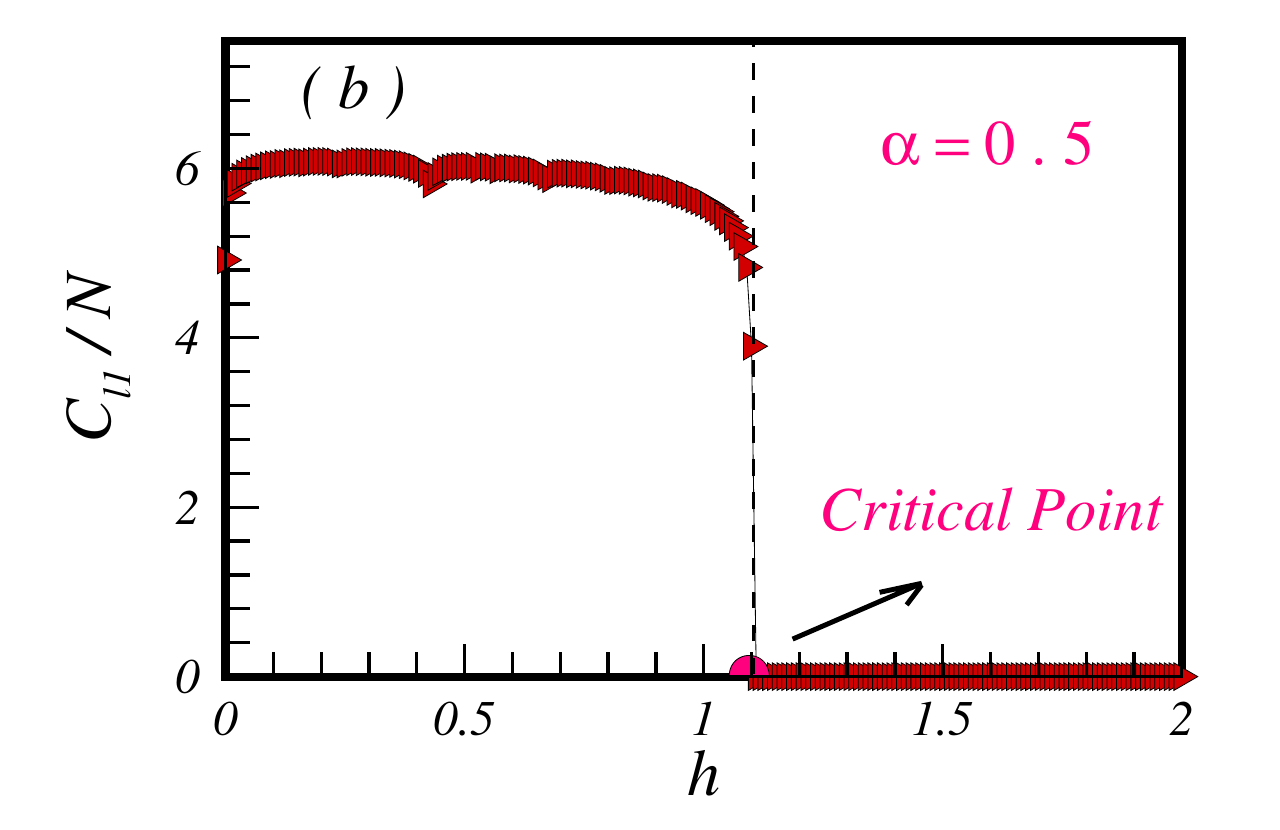}  }
\centerline{\includegraphics[width=0.55\linewidth]{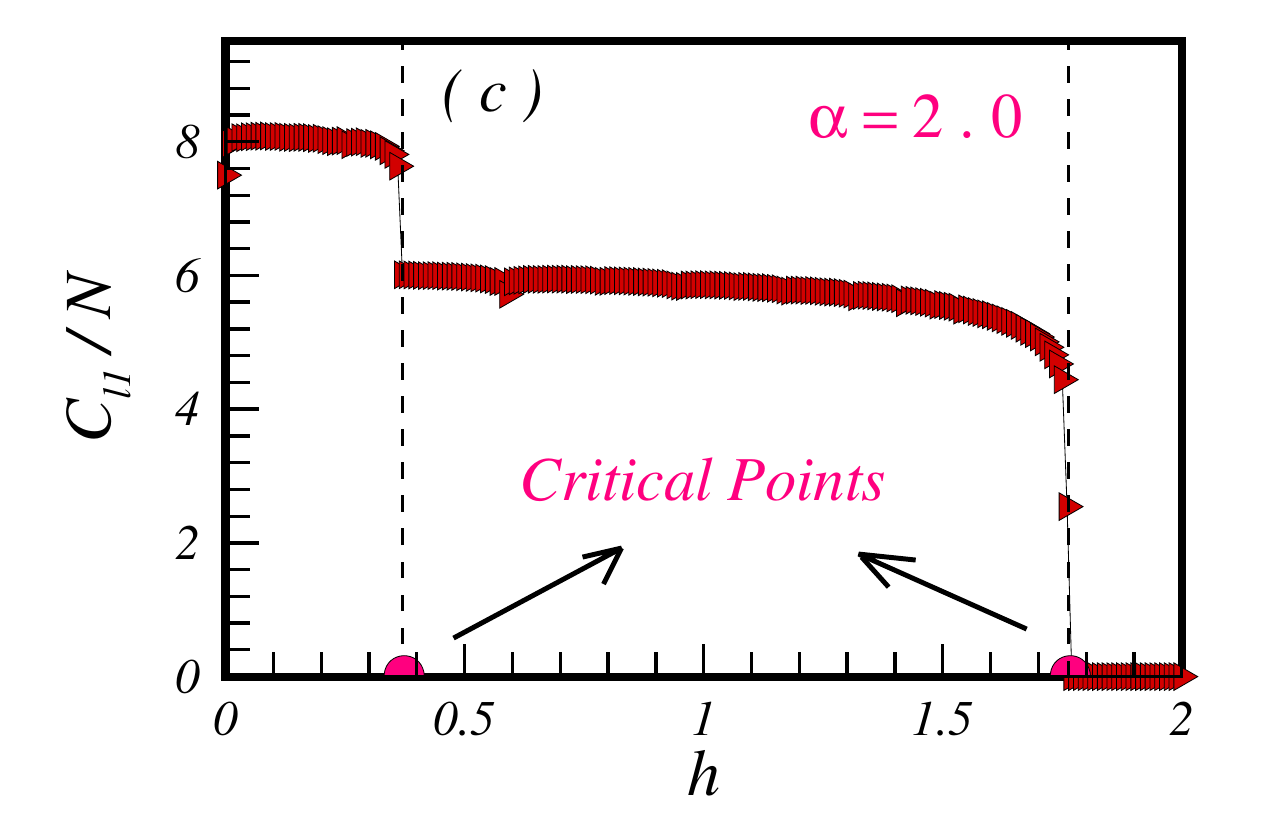} }
\caption{(color online).  The $l_1$-norm of coherence with respect to the TSI and the magnetic field. Results are presented for a chain size $N=1000$. (a) The effect of the TSI in absence of the magnetic field $h=0$. (b) The effect of the magnetic field when the system is  first located in the SL-I phase by selecting $\alpha=0.5$. (c) The effect of the magnetic field when the system is first  located in the SL-II phase by selecting $\alpha=2.0$.  }
\label{Fig1}
\end{figure}

\section{Results}

We present our results in three sections. First, we analyze the coherence of the ground state, identifying regions of the phase diagram with coherence and incoherence. Second, we explore the SSP to find unique coherent states with uniform noise. Third, we investigate the EE to identify regions of entanglement when the system is divided into two equal parts.

\begin{figure}
\centerline{\includegraphics[width=0.55\linewidth]{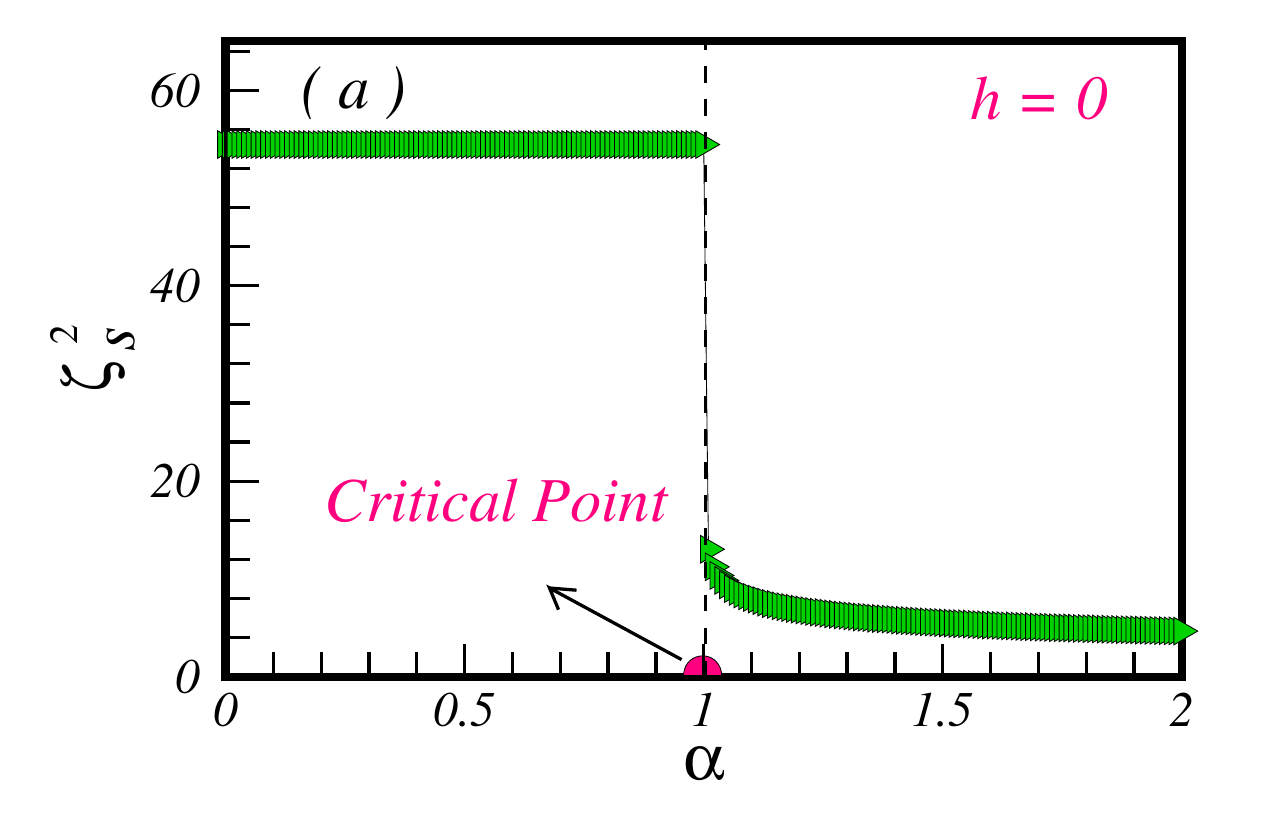}  \includegraphics[width=0.55\linewidth]{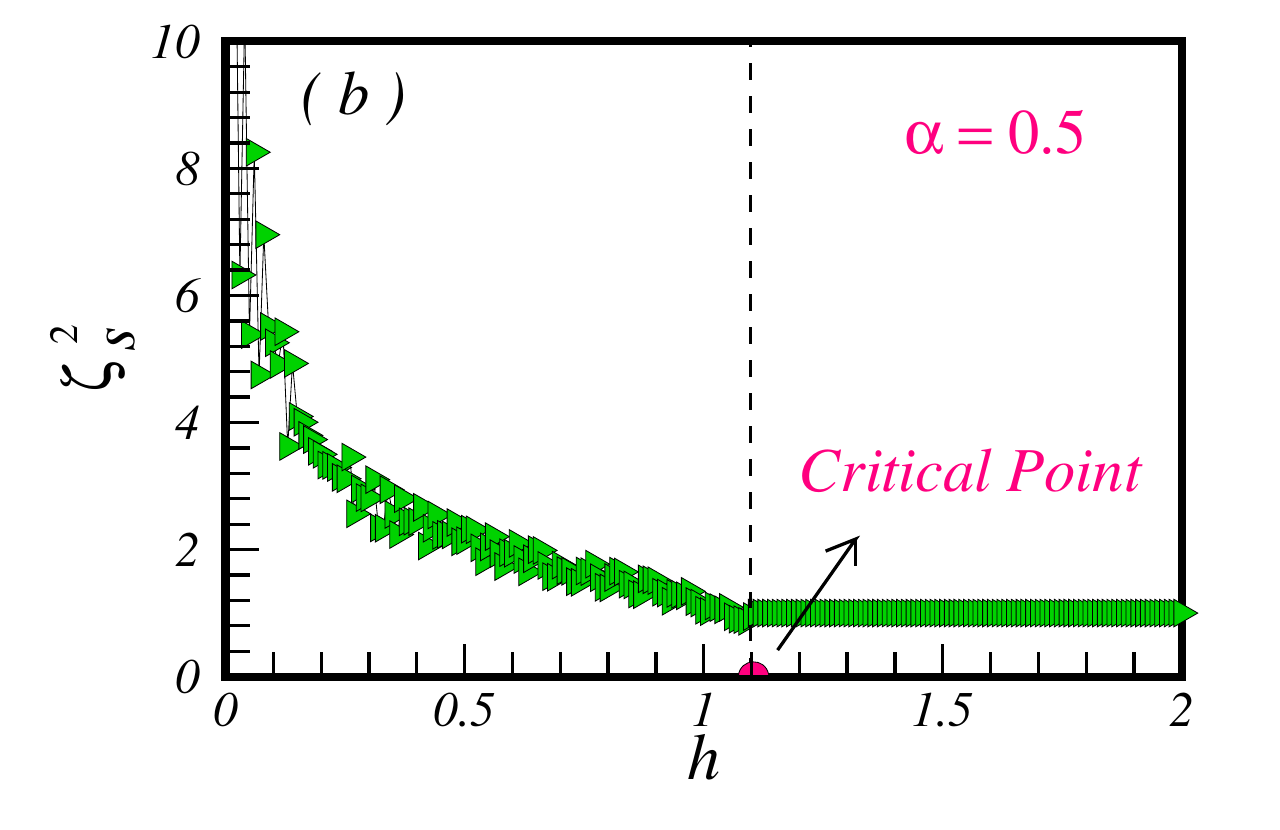}  }
\centerline{\includegraphics[width=0.55\linewidth]{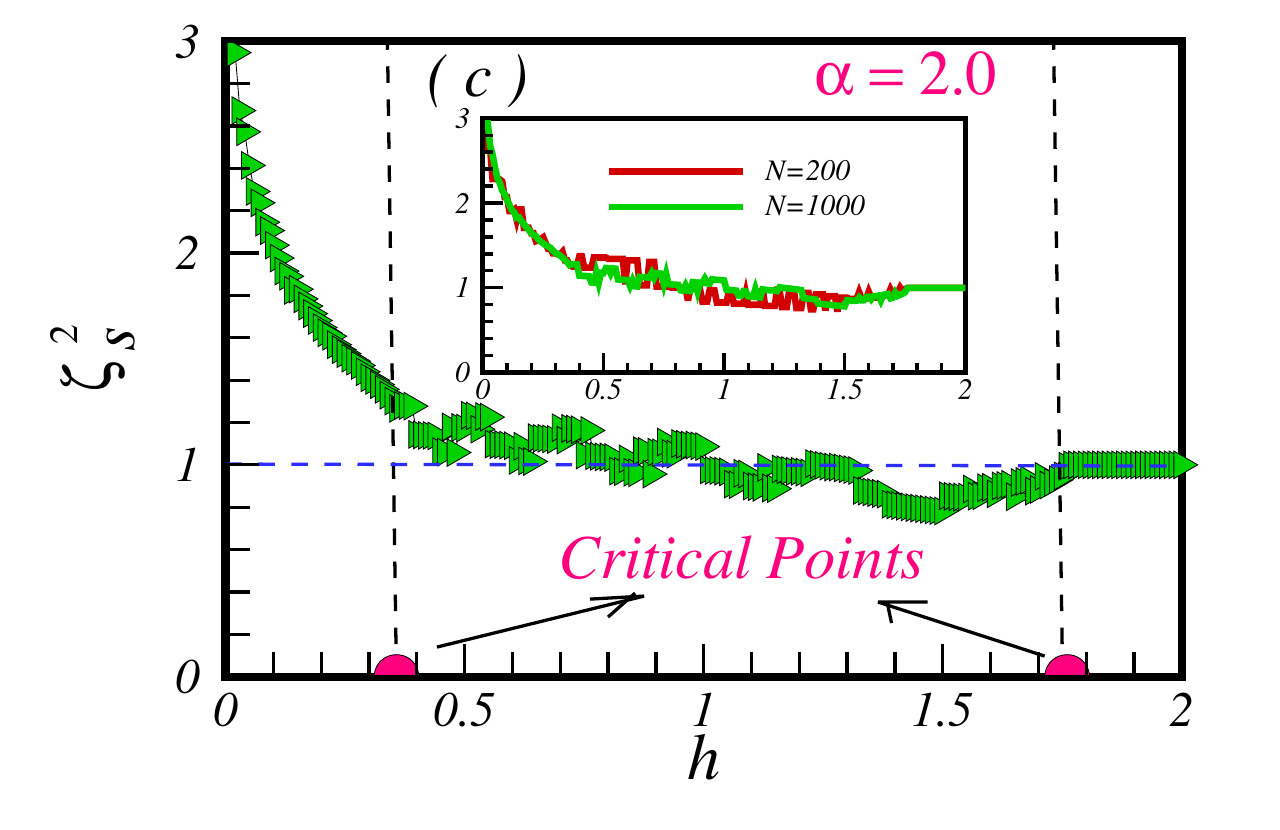} }
\caption{(color online). The SSP with respect to the TSI and the magnetic field. Results are presented for a chain size $N=1000$. (a) The effect of the TSI in absence of the magnetic field $h=0$. (b) The effect of the magnetic field when the system is first located in the SL-I phase by selecting $\alpha=0.5$. (c) The effect of the magnetic field when the system is first located in the SL-II phase by selecting $\alpha=2.0$. The inset displays the results for two chain sizes $N=200, 1000$, where increasing the system size reduces the fluctuations significantly.}  
\label{Fig2}
\end{figure}

In Fig.~\ref{Fig1}, we present the results of the scaled $l_1$-norm of coherence ($C_{l_1}/N$). In Fig.~\ref{Fig1} (a), we observe that the ground state of the spin-1/2 XX chain model exhibits coherence when no magnetic field is applied, and the cluster TSI has no discernible effect on the $l_1$-norm function in the SL-I phase. This phenomenon suggests a novel concept termed the 'gapless plateau state,' where the $l_1$-norm of coherence remains constant, coinciding with an energy gap of zero over a finite TSI range.  Furthermore, this observation can be attributed to the fact that the spin-spin correlations remain independent of the TSI in the SL-I phase  \cite{Ne-2}.  At the quantum critical point ($\alpha_c=1.0$), the $l_1$-norm function experiences a discontinuity, jumping to a higher value, indicative of a phase transition. In the SL-II phase, the system's ground state displays increased coherence with fluctuations. Fig.~\ref{Fig1} (b)-(c) illustrate the influence of the magnetic field on the ground state. In the SL-I phase (Fig.~\ref{Fig1} (b)), coherence gradually diminishes with an increasing magnetic field until reaching zero at the critical field. In the PM phase, the ground state becomes incoherent. In the SL-II phase, the behavior of ground state coherence differs, as shown in Fig.~\ref{Fig1} (c). The scaled $l_1$-norm of coherence exhibits drops at both critical fields. At the first critical field, it decreases to a lower level of coherence, while at the second critical field, it drops to zero, signifying incoherence in the PM phase.


In Fig.~\ref{Fig2}, we depict the variation of the SSP concerning the TSI and the external magnetic field. In  Fig.~\ref{Fig2} (a), we observe that the SSP remains constant within the SL-I phase and gradually diminishes as the TSI interaction increases. Notably, the ground state exhibits non-squeezing behavior in the SL-I phase, reflecting what we term a 'gapless plateau state.' The SSP experiences a sharp decline at the critical point $\alpha_c=1$, signifying a quantum phase transition into the gapless SL-II phase. Within the SL-II phase, the ground state similarly exhibits non-squeezed characteristics. Fig.~\ref{Fig2} (b)-(c) elucidate how the SSP behaves in the presence of a magnetic field. In the SL-I phase, the SSP exhibits fluctuations with increasing field strength but remains in a non-squeezed state until it reaches the critical field $h_c(\alpha)$. At this point, the system transitions into a fully polarized and coherent state. Within the SL-II phase, the SSP steadily decreases, reaching a value slightly above the first critical field $h_{c_1}$. This transition indicates that the system becomes coherent according to the SSP criterion. Subsequently, the SSP alternates between squeezed and non-squeezed states until it reaches the saturation critical field, signifying a return to a coherent state. The inset in Fig.~\ref{Fig2} (c) illustrates the SSP for spin chains with sizes $N = 200, 1000$ and $\alpha=2.0$. Increasing the system size reduces fluctuations, indicating their origin in numerical finite size effects.  

\begin{figure}
\centerline{\includegraphics[width=0.65\linewidth]{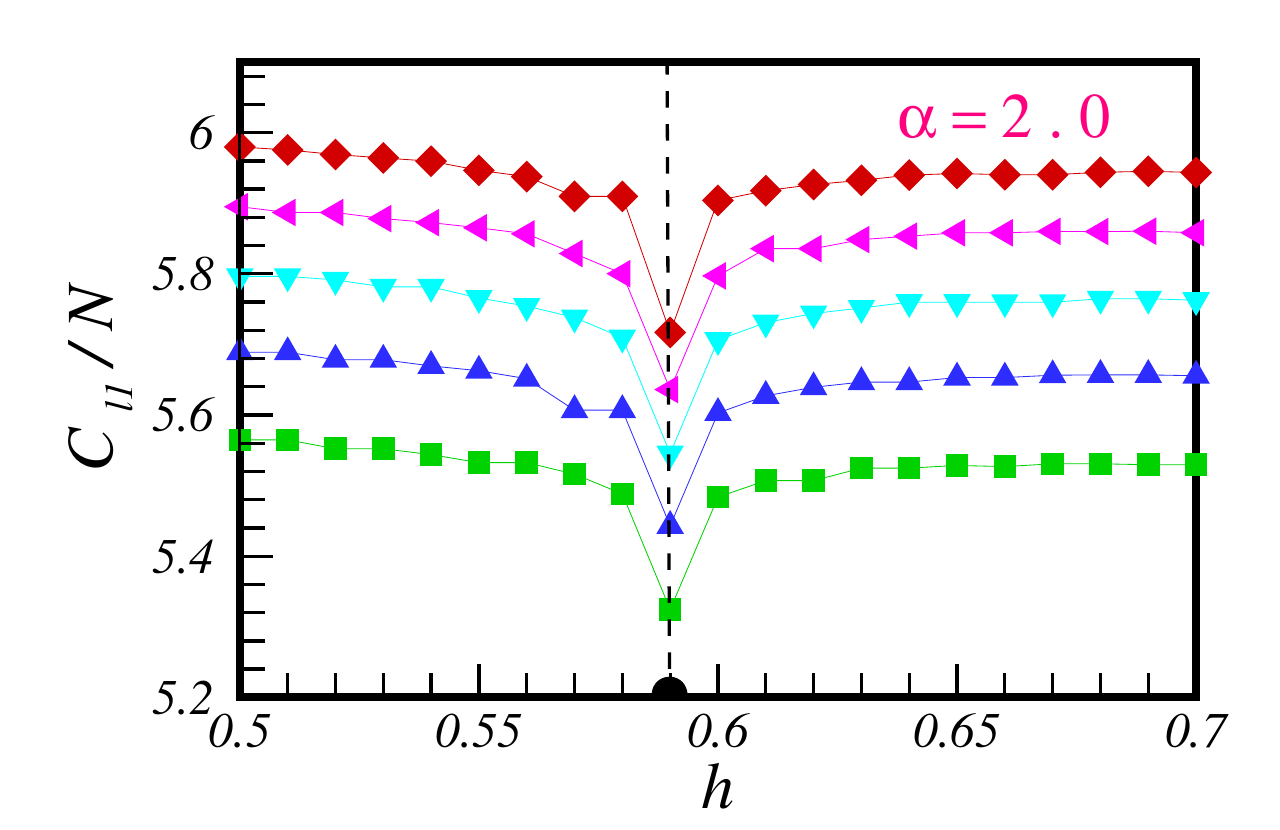}  }
\caption{(color online). The $l_1$-norm of coherence with respect to the magnetic field. The TSI is selected at $\alpha=2.0$. Results are presented for  chain sizes $N=600, 700, 800, 900,$ and $1000$ from bottom to top.  }
\label{Fig2-0}
\end{figure}

As noted, when $\xi _s^2=1$, the ground state represents a unique coherent state with equal noise in magnetization across all directions. Therefore, we reexamined the $l_1$-norm of coherence to identify these distinctive coherent states. Results are presented in Fig.~\ref{Fig2-0} for various chain sizes $N=600, 700, ... , $ and $1000$ at $\alpha=2.0$. Surprisingly, we observe a minimum in the $l_1$-norm of coherence at a precise magnetic field value, $h_{coh}=0.59$. Consequently, we infer that the $l_1$-norm of coherence can effectively detect these special coherent states with $\xi _s^2=1$, as they correspond to local minima in the coherence measurement.

\begin{figure}
\centerline{\includegraphics[width=0.55\linewidth]{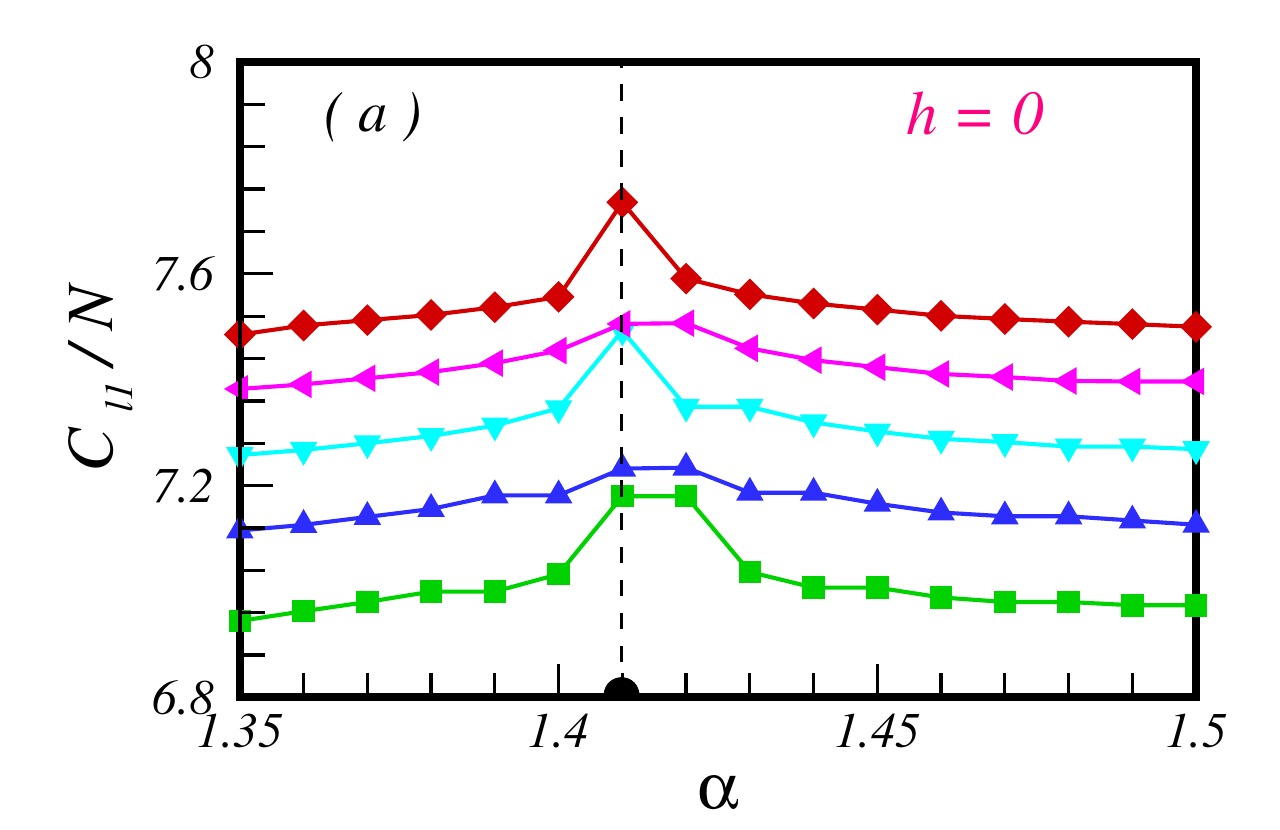}  \includegraphics[width=0.55\linewidth]{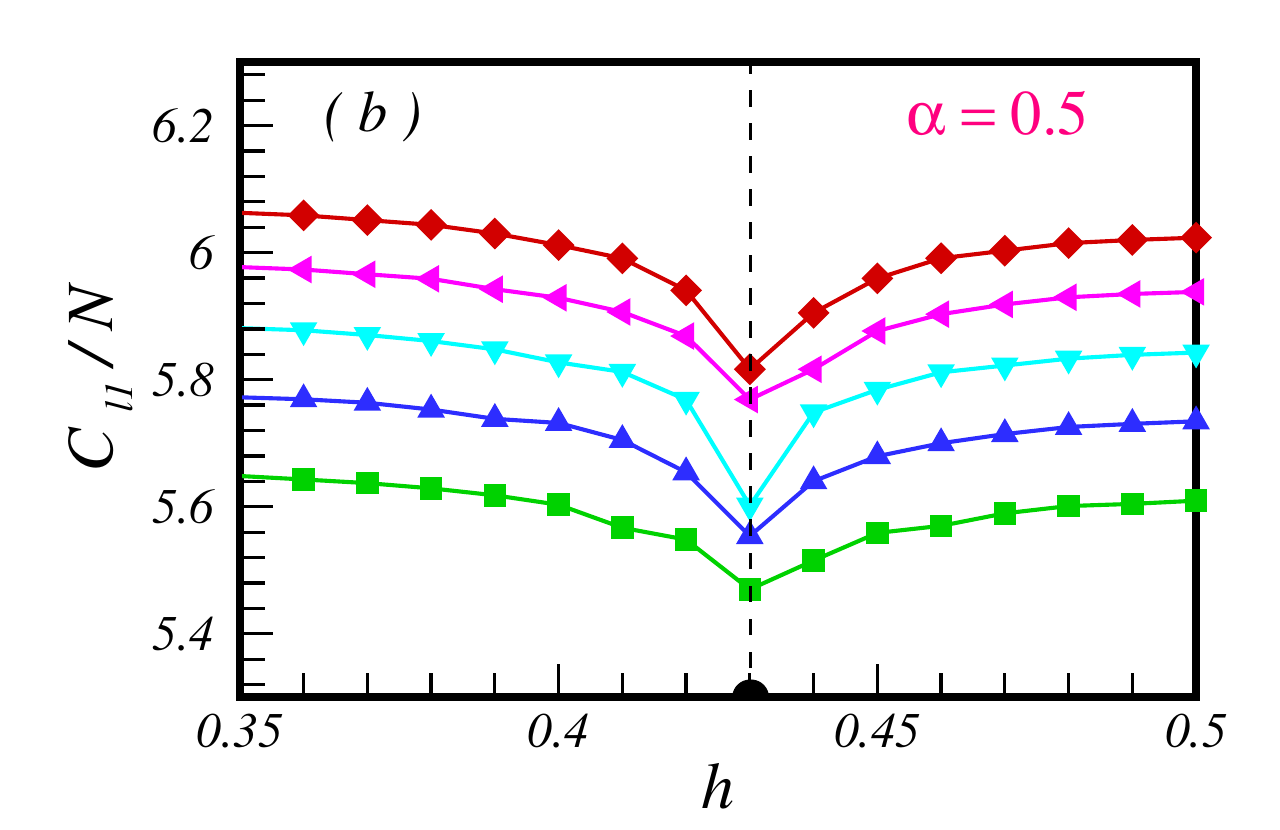}  }
\centerline{\includegraphics[width=0.55\linewidth]{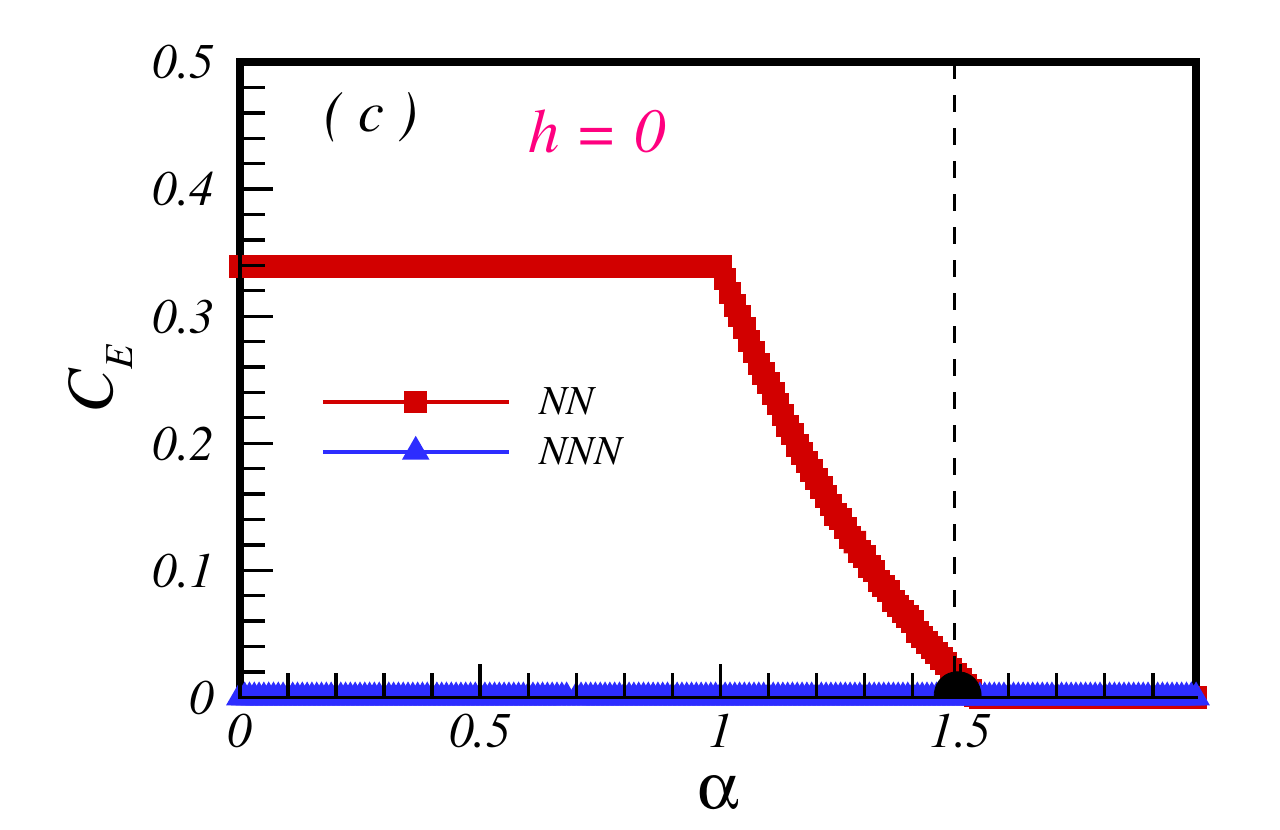}
\includegraphics[width=0.55\linewidth]{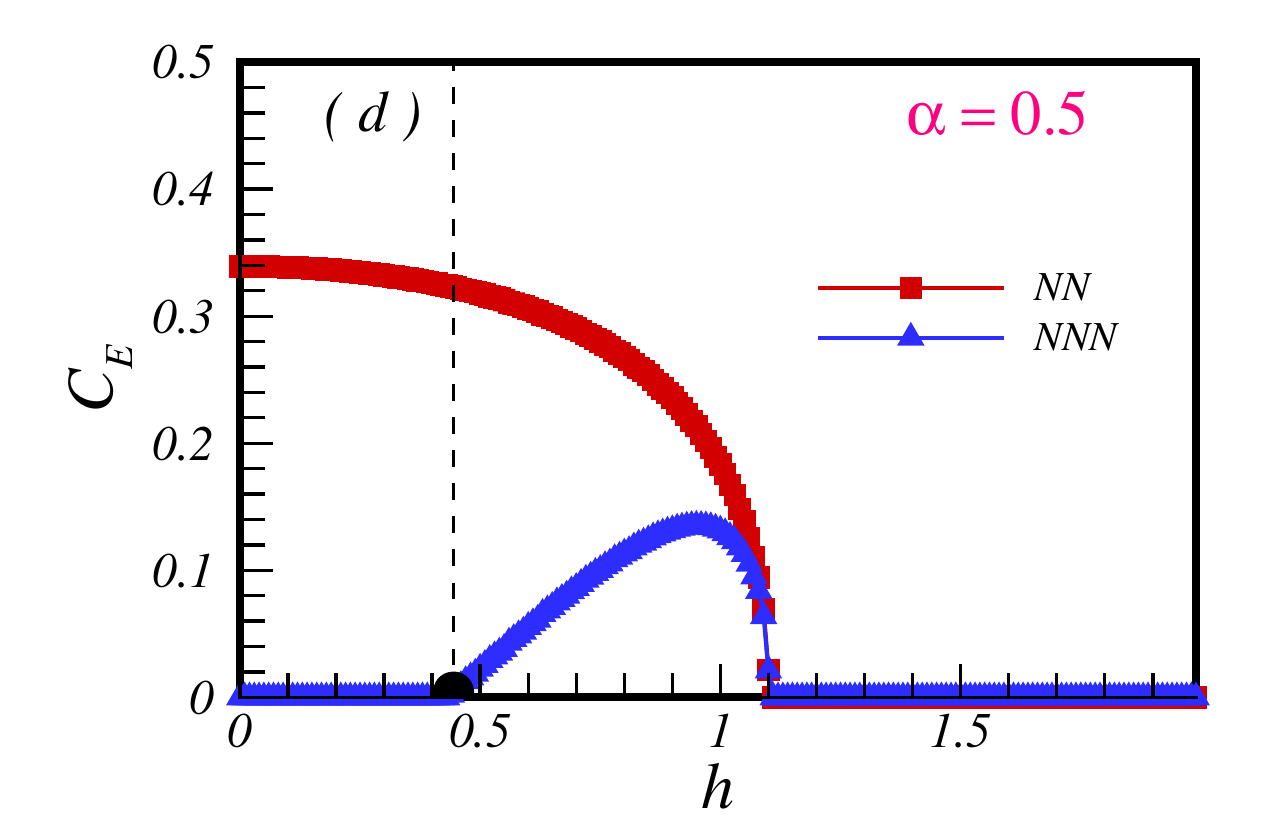} }
\caption{(color online).  The $l_1$-norm of coherence is presented with respect to (a) the TSI in the absence of the magnetic field and (b) the magnetic field for a certain value of TSI ($\alpha=0.5$). Results are shown for chain sizes $N = 600, 700, 800, 900$ and $1000$ from bottom to top. The concurrence between the nearest neighbor and next-nearest-neighbor pairs of spins is depicted as a function of (c) TSI in the absence of the magnetic field and (d) the magnetic field for $\alpha=0.5$. }    
\label{Fig3-0}
\end{figure}

We observed irregularities in the curves of the $l_1$-norm of coherence, particularly  around $\alpha=1.5$ without the TF and around $h=0.5$ at TSI $\alpha=0.5$, where the system starts in the SL-I phase. To verify whether these are numerical artefacts, we compute the $l_1$-norm of coherence for various chain sizes $N=600, 700, ... , $ and $1000$, and presented the results in Fig.~\ref{Fig3-0} (a) and (b). Evidently, the $l_1$-norm of coherence exhibits extrema around these points, confirming that they are not due to finite-size effects. Since the $l_1$-norm is evaluated on the ground state of the system, we investigated the physical origin of these signals. Our focus turned to the entanglement between two spins, quantified by the concurrence. The concurrence, a measure of entanglement, captures  quantum correlation between two spins in the system \cite{con-1,con-2,con-3}. In recent years, the concurrence and other bipartite quantum correlations have proven valuable in solving complex problems related to the ground state of spin-1/2 low dimensional systems \cite{con-4,con-5,con-6, con-7, con-8}.

For two arbitrary spins at position $i$ and $j$, the two-site
reduced density matrix generally takes the form,

\begin{eqnarray}\label{eq01}
\rho_{i,j}&=&\frac{1}{4}+\sum\limits_{\alpha} (\langle S_{i}^{\alpha} \rangle S_{i}^{\alpha}+\langle S_{j}^{\alpha} \rangle S_{j}^{\alpha})\\ \nonumber
&+&\sum\limits_{\alpha, \beta} \langle S_{i}^{\alpha}  S_{j}^{\beta} \rangle S_{i}^{\alpha}  S_{j}^{\beta},
\end{eqnarray}
where $\alpha, \beta=x, y, z$. The concurrence between two spin-1/2 particle at sites $i$ and $j$ can be obtained  from the corresponding reduced density matrix $\rho_{ij}$.  The reduced density matrix in the standard basis ($|\uparrow\uparrow\rangle, |\uparrow\downarrow\rangle, |\downarrow\uparrow\rangle, |\downarrow\downarrow\rangle $) is expressed as

\begin{equation}\label{eq6}
{\rho _{i,j}} = \left( {\begin{array}{*{20}{c}}
{{\langle p_{i}^{\uparrow} p_{j}^{\uparrow} \rangle}}&  {{\langle p_{i}^{\uparrow} S_j^{-} \rangle}}  &{{\langle S_i^{-} p_{j}^{\uparrow}  \rangle}}&{{\langle S_i^{-} S_j^{-}  \rangle}}\\
{{\langle p_{i}^{\uparrow} S_j^{+} \rangle}}&{{\langle p_{i}^{\uparrow} p_{j}^{\downarrow} \rangle}}&{{\langle S_i^{-} S_j^{+} \rangle}}&{{\langle S_i^{-} p_{j}^{\downarrow}  \rangle}}\\
{{\langle S_i^{+} p_{j}^{\uparrow}  \rangle}}&{{\langle S_i^{+} S_j^{-}  \rangle}}&{{\langle p_{i}^{\downarrow} p_{j}^{\uparrow} \rangle}}&{{\langle p_{i}^{\downarrow} S_j^{-} \rangle}}\\
{{\langle S_i^{+} S_j^{+}  \rangle}}&{{\langle S_i^{+} p_{j}^{\downarrow}  \rangle}}&{{\langle p_{i}^{\downarrow} S_j^{+}   \rangle}}&{{\langle p_{i}^{\downarrow} p_{j}^{\downarrow} \rangle}}
\end{array}} \right),
\end{equation}
where the brackets denote the average over the physical state and $p^{ \uparrow  }=\frac{1}{2}+S^{z}$,  $p^{\downarrow}=\frac{1}{2}-S^{z}$, $S^{\pm}=S^{x}\pm i S^{y}$. The concurrence of two spins is defined as $C_{E}=\max (0,\lambda_1-\lambda_2-\lambda_3-\lambda_4$), where $\lambda_{i}$ is the square root of the eigenvalue of $R=\rho_{i,j} \tilde{\rho}_{i,j}$ and $\tilde{\rho}_{i,j}=(\sigma_i^{y} \otimes \sigma_j^{y}) \rho_{i,j}^{\star} (\sigma_i^{y} \otimes \sigma_j^{y}) $. 
Due to the symmetry of the Hamiltonian, most of the off-diagonal elements of the reduced density matrix $\rho_{i,j}$ are zero. First, the translational symmetry implies that the density matrix satisfies $\rho_{i,j}=\rho_{i,i+r}$ for any position $i$. Second, the spin-1/2 XX chain model with TSI in a magnetic field has a $Z_2$ symmetry, that it is invariant under $\pi$-rotation around the $z$-axis. This also implies that the density matrix commutes with the operator $S_{i}^{z}  S_{j}^{z}$. Based on these symmetry properties, the density matrix must be symmetric, and only some elements of the reduced density matrix are non-zero,

%

\begin{equation}\label{eq6}
{\rho _{i,j}} = \left( {\begin{array}{*{20}{c}}
{{X_{i,j}^ +}}&0&0&0\\
0&{{Y_{i,j}^ +}}&{{Z_{i,j}^*}}&0\\
0&{{Z_{i,j}}}&{{Y_{i,j}^ -}}&0\\
0&0&0&{{X_{i,j}^ -}}
\end{array}} \right).
\end{equation}
Finally, the concurrence is given by the following expression
\begin{eqnarray}
C_{E}&=& \max \{0, 2 (|Z_{i,j}|-\sqrt{X_{i,j}^{+} X_{i,j}^{-}}) \}. 
\label{Concurr}
\end{eqnarray}

We obtain analytical results for the concurrence of nearest-neighbor and next-nearest neighbor pair of spins for a chain size $N=1000$, shown in Fig.~\ref{Fig3-0} (c) and (d). In Fig.~\ref{Fig3-0} (c),  the next-nearest neighbor pair of spins are unentangled without the magnetic field, and the TSI does not create entanglement between them. However, nearest neighbor pair of spins are entangled without the magnetic field and they stay entangled in the SL-I phase as the TSI increases. At the critical TSI, $\alpha_c=1.0$, the concurrence of nearest neighbor pair of spins begins to drop and it disappears around $\alpha=1.5$ as the TSI increases further. This explains the anomaly in the $l_1$-norm of coherence at that point, indicating a specific value of the TSI where the nearest neighbor pair of spins become unentangled. We perform the same calculation for $\alpha=0.5$ and show the results in Fig.~\ref{Fig3-0} (d). As we see, initially, nearest neighbor pair of spins are entangled and by increasing the magnetic field, entanglement between them reduces and becomes zero at the saturation field. In the paramagnetic phase, there is no entanglement as expected. Interestingly, next nearest neighbor pair of spins gets entangled at a certain value of the magnetic field which is the same as the anomaly point (around $h=0.5$) in the $l_1$-norm of coherence.

\begin{figure}
\centerline{\includegraphics[width=0.55\linewidth]{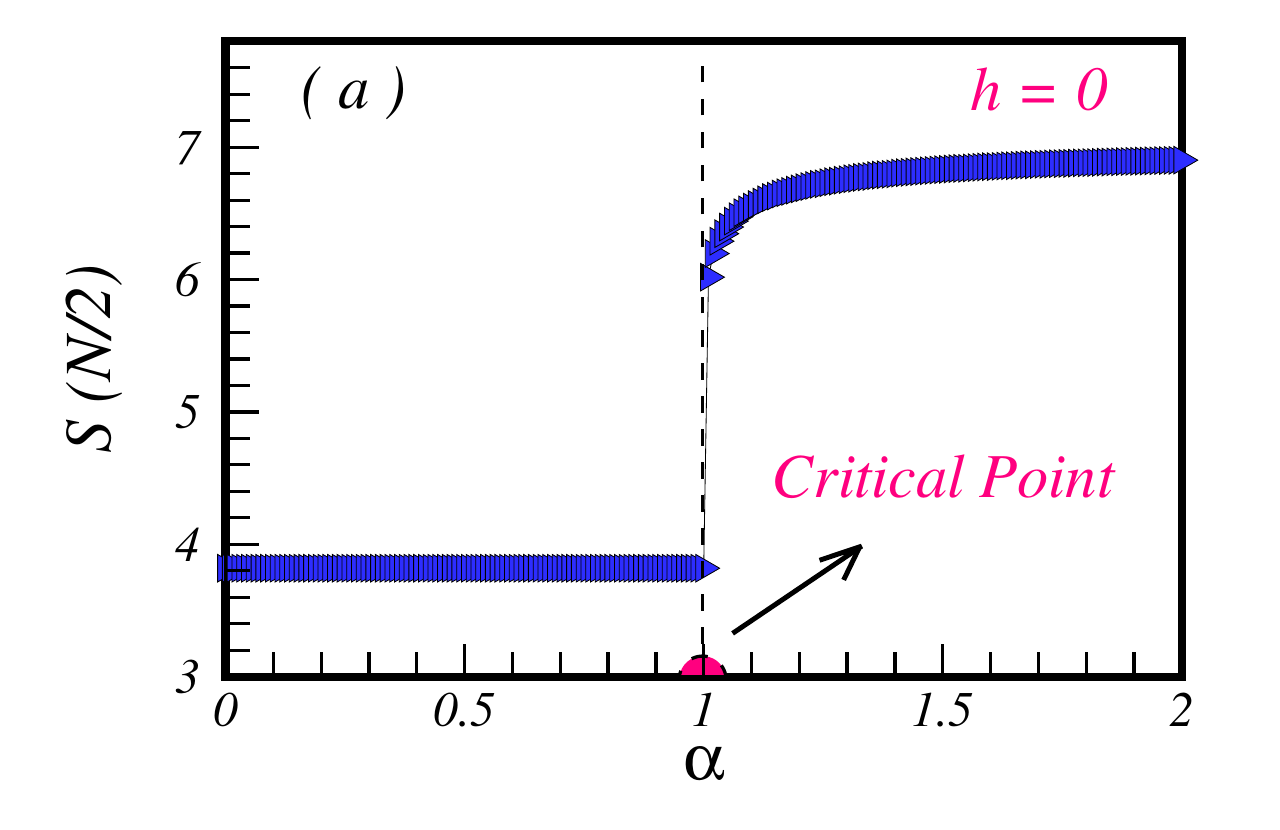}  \includegraphics[width=0.55\linewidth]{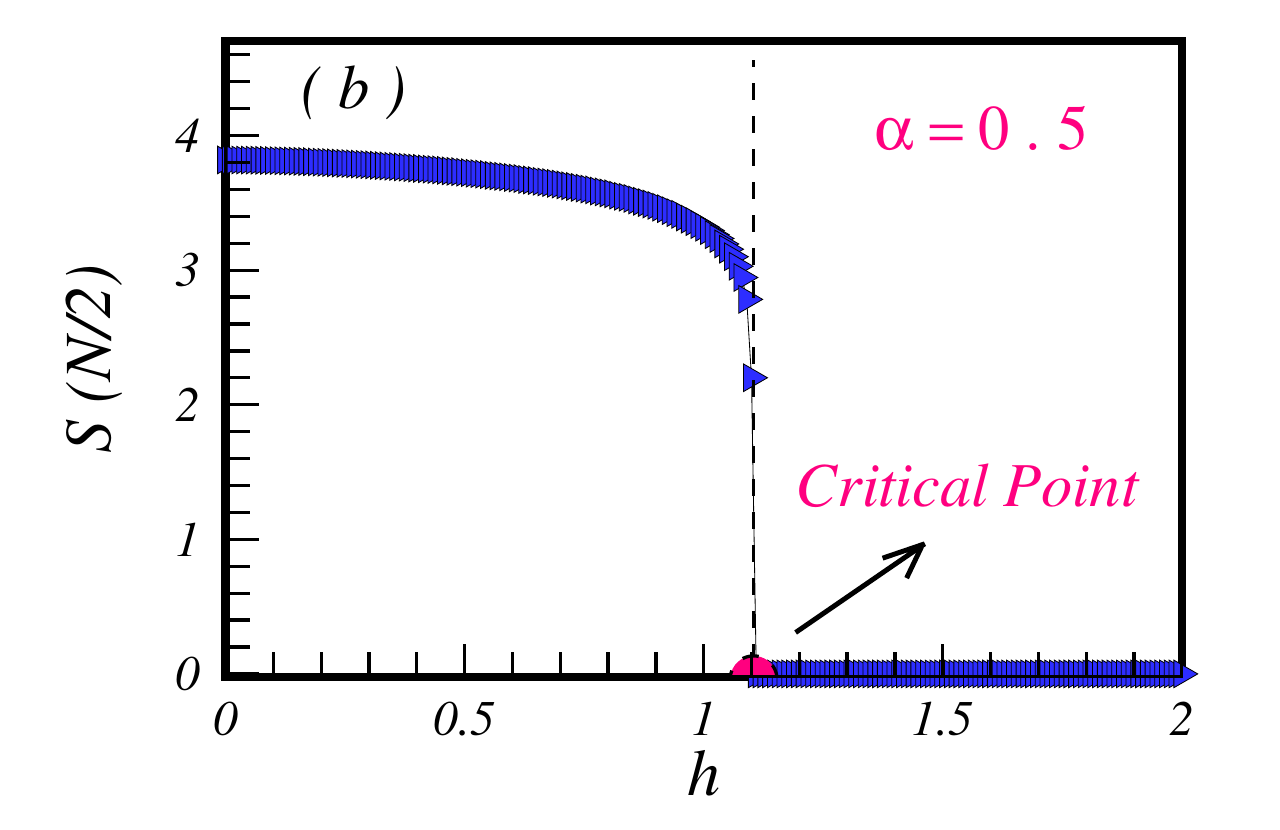}  }
\centerline{\includegraphics[width=0.55\linewidth]{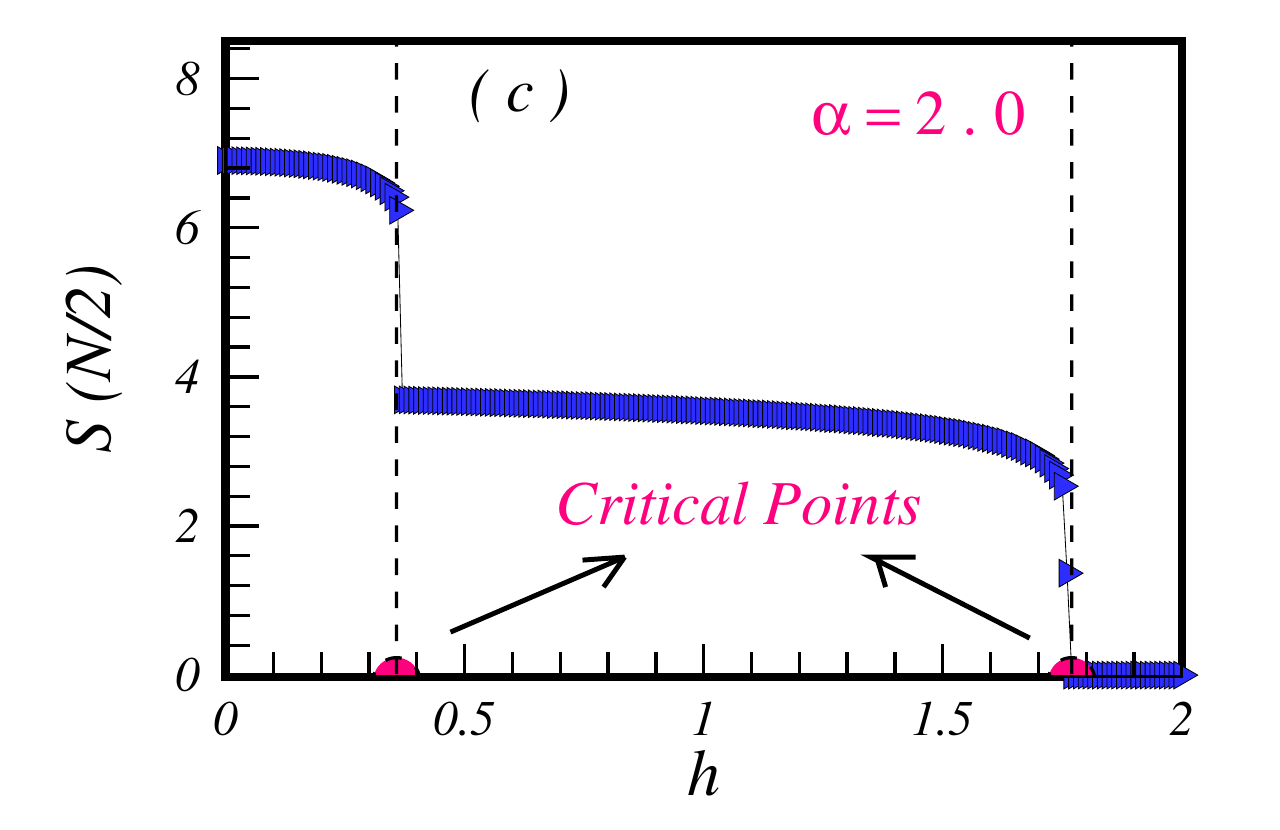} }
\caption{(color online). The EE with respect to the TSI and the magnetic field. Results are presented for a chain size $N=1000$. (a) The effect of the TSI in the absence of the magnetic field $h=0$. (b) The effect of the magnetic field when the system is first located in the SL-I phase by selecting $\alpha=0.5$. (c) The effect of the magnetic field when the system first  is located in the SL-II phase by selecting $\alpha=2.0$.  }
\label{Fig3}
\end{figure}

We conducted an analysis of the EE between two equal parts of the system across various values of the TSI and the magnetic field. EE can provide crucial insights into the system's quantum phases, transitions, and information processing capabilities. Our results, illustrated in  Fig.~\ref{Fig3}, reveal several key observations. 
In Fig.~\ref{Fig3}, we observe that EE exhibits a distinct behavior compared to the $C_{l_1}$ and SSP metrics. Unlike the fluctuations noted in $C_{l_1}$ and SSP, EE remains relatively stable. Notably, we find that the two parts of the system exhibit entanglement when there is no magnetic field, particularly in the SL-I phase. This entanglement is unaffected by the TSI (Fig.~\ref{Fig3} (a)). However, at the critical point $\alpha_c=1.0$, we observe a marked increase in entanglement. In the SL-II phase, the two parts display higher correlations than in the SL-I phase. Furthermore, within the SL-II phase, entanglement increases with higher TSI values. Our results indicate that the ground state of the system in the SL-II phase exhibits higher quantum correlations or entanglement compared to the SL-I phase. This suggests that the SL-II phase possesses greater conformal symmetry or complexity than the other region. The disparity in the EE can be attributed to differences in the central charge or the correlation length exponent of the distinct phases. It is important to note that the magnetic field exerts a reduction effect on the entanglement between the two parts, irrespective of whether the system is in the SL-I or SL-II phases (Fig.~\ref{Fig3} (b)-(c)). At all critical fields, we observe a decrease in entanglement. Notably, the PM phase exhibits no entanglement between the two system parts.
These findings shed light on the intricate relationship between entanglement and system parameters, providing valuable insights into the quantum behaviors of the system.

Scaling behavior is a fundamental phenomenon that reflects the self-similarity and universality of a system in the vicinity of a critical point, where the system undergoes a phase transition. This behavior is characterized by the emergence of scaling laws and scaling functions, which incorporate essential parameters such as scaling exponents and scaling variables. These scaling laws and functions play a crucial role in describing how physical properties of a system change as it approaches a critical point. They offer insights into the critical behavior of the system, highlighting the common features and patterns that are independent of specific details, making them valuable tools in the study of phase transitions and critical phenomena.

In the following sections, we investigate how the $l_1$-norm of coherence, the SSP, and the EE functions vary with changes in the system size. Analyzing the scaling behavior of these quantities is of paramount importance as it unveils the universal characteristics associated with quantum phase transitions and quantum criticality. The practical utility of these scaling analyses in quantitatively characterizing quantum phase transitions is an exciting and ongoing area of research.

\begin{figure}
\centerline{\includegraphics[width=0.55\linewidth]{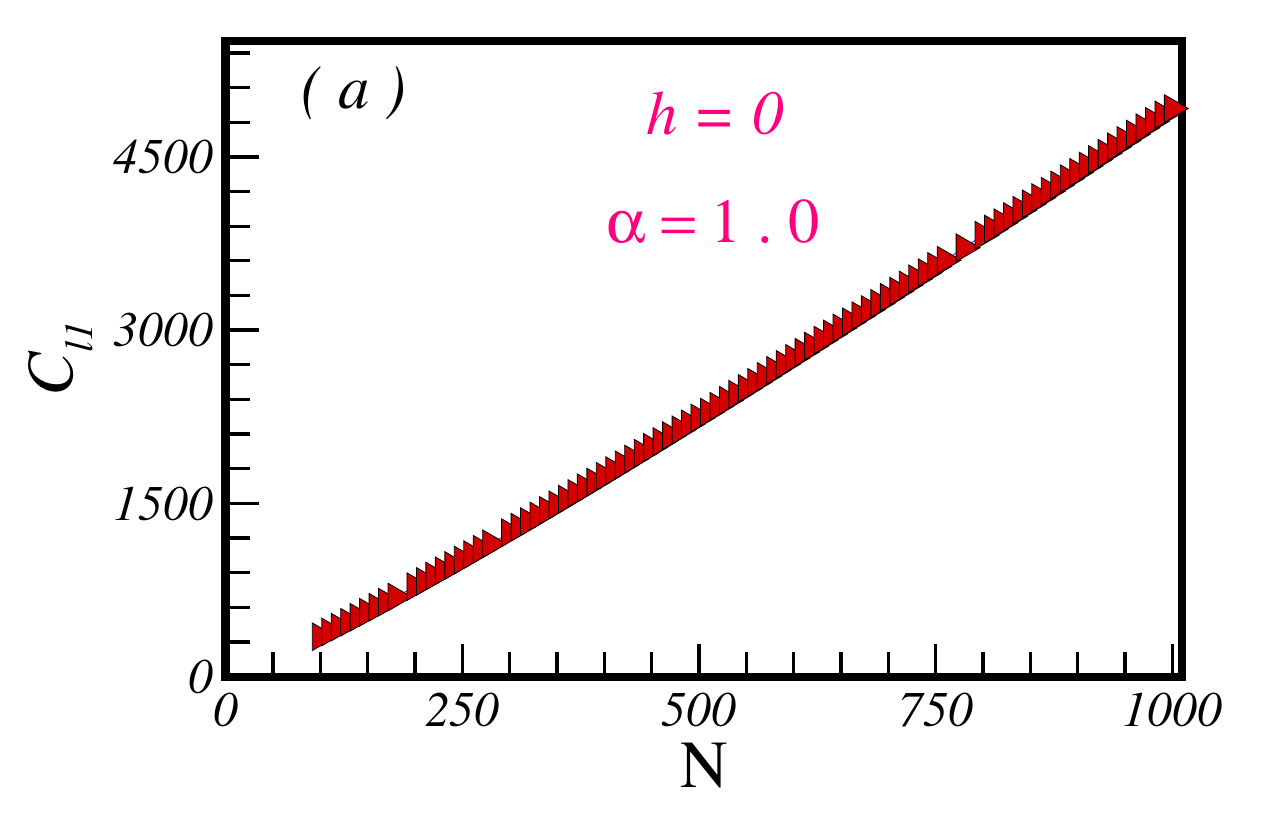}  \includegraphics[width=0.55\linewidth]{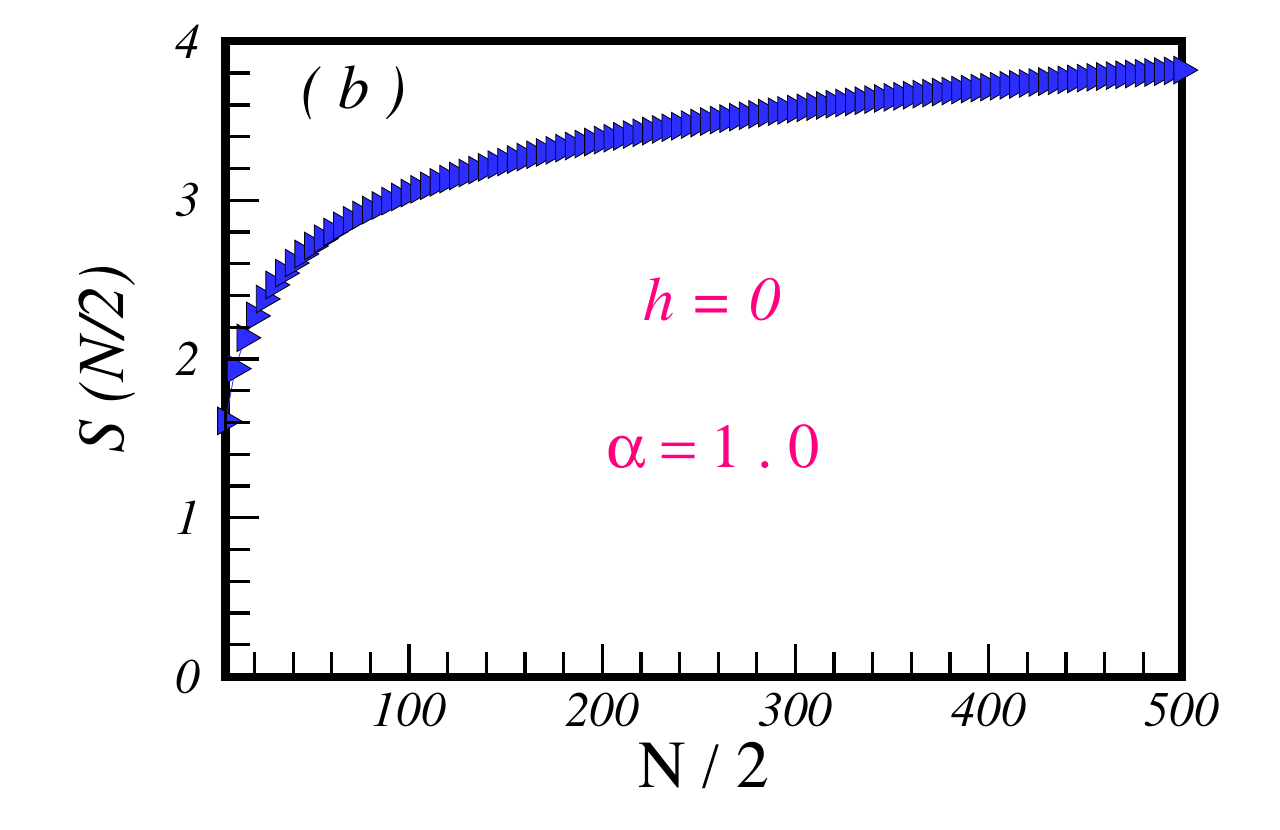}  }
\centerline{\includegraphics[width=0.55\linewidth]{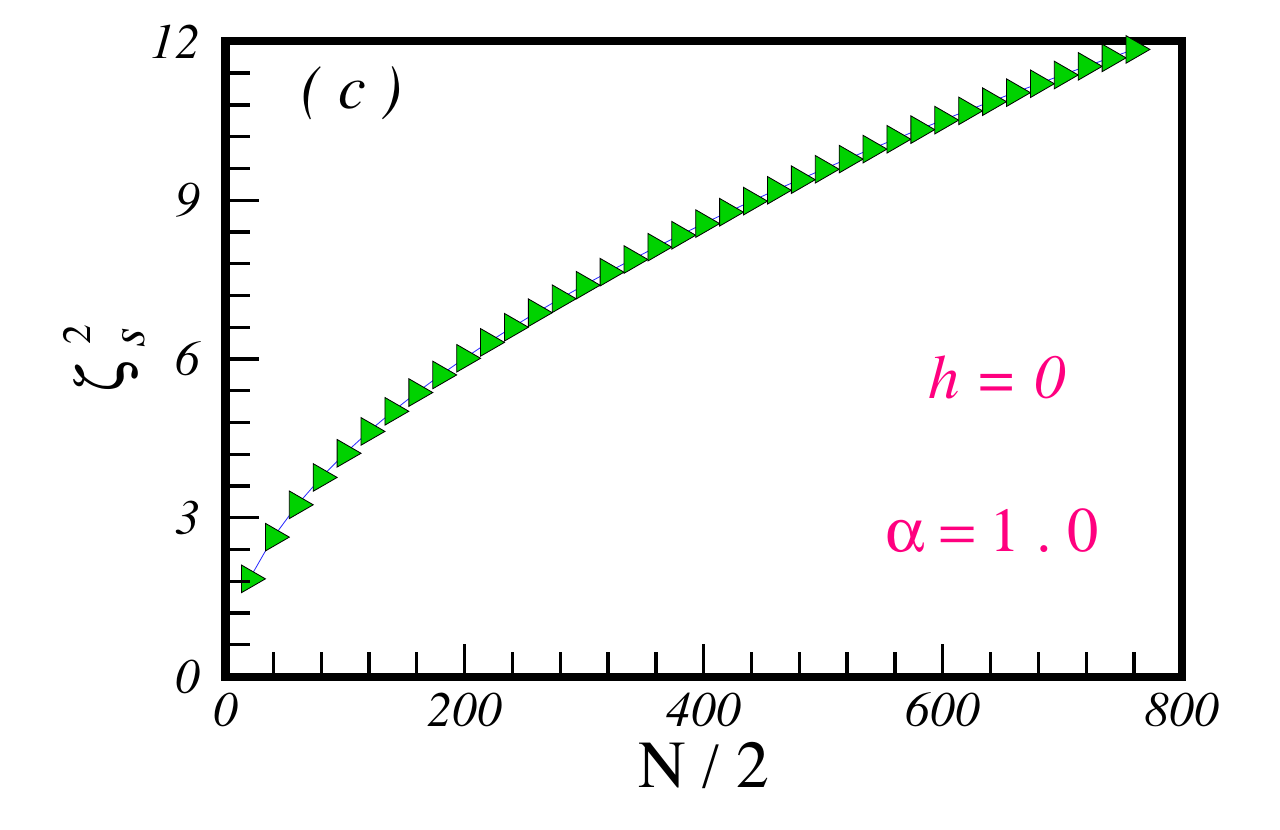} }
\caption{(color online). Scaling behaviour with respect to the system size. Results are calculated at $h=0$ and $\alpha=\alpha_c=1$, a critical point separating  gapless SL-I and SL-II phases.  (a) the $l_1$-norm of coherence, (b) the EE and (c) the SSP.  }
\label{Fig4}
\end{figure}

Scaling of EE finds significant applications in various domains, particularly within condensed matter physics, quantum information theory, and quantum computation \cite{SC-1,SC-2,SC-3, SC-3-0}. In condensed matter physics, it serves as a valuable tool for classifying and characterizing different phases of matter, including topological phases, spin liquids, and symmetry-protected topological phases. In quantum information theory, scaling of EE is instrumental in quantifying the complexity and information content of quantum states, such as matrix product states, tensor network states, and holographic states. Moreover, in quantum computation, the scaling of EE plays a role in designing and optimizing quantum algorithms, spanning applications in quantum simulation, quantum annealing, and quantum machine learning.

The applications of SSP scaling primarily benefit the fields of quantum metrology and quantum sensing. Within these domains, two fundamental limits govern the precision of quantum measurements in spin systems: the standard limit and the Heisenberg limit. 
The standard limit, also known as the shot-noise limit or standard quantum limit, represents the precision achievable using uncorrelated or coherent spin states \cite{SC-4,SC-5}. It scales as $1/\sqrt{N}$, with $N$ being the number of spins. 
In contrast, the Heisenberg limit, also referred to as the ultimate limit or quantum Cramér-Rao bound, embodies the precision attainable when employing entangled or squeezed spin states. It scales as $1/N$ \cite{SC-6,SC-7,SC-8}. Remarkably, the Heisenberg limit sets the highest possible precision achievable by any quantum state and outperforms the standard limit by a factor of $1/\sqrt{N}$. 
The concept of spin squeezing, which reduces quantum fluctuations in one spin component below the standard limit through quantum correlations among spins, holds the potential to enhance measurement precision beyond the standard limit. It's important to note, however, that not all spin-squeezed states can reach the Heisenberg limit. Only specific spin-squeezed states, such as the two-axis twisting states or the Dicke states, can achieve the Heisenberg limit under specific conditions \cite{SC-9}. 
Consequently, the study of spin squeezing with various interactions, including TSI, holds significant relevance in the fields of quantum metrology and quantum information science.

We present our scaling results in  Fig.~\ref{Fig4}, focusing on the $l_1$-norm of coherence, the SSP and the EE of the ground state in the XX chain model with TSI interaction at the quantum critical point ($\alpha_c=1$). 
Fig.~\ref{Fig4} (a) illustrates that the $l_1$-norm of coherence exhibits linear scaling with system size, consistent with the critical exponent of the correlation length. This scaling behavior remains consistent across different magnetic phases and aligns with predictions from renormalization group theory \cite{SC-10}.

In Fig.~\ref{Fig4} (b) a logarithmic divergence of the EE is observed as the system size increases, signaling a departure from the area law. This divergence indicates that the EE captures the universal properties of the critical point and the system's number of degrees of freedom. 
It is known that for gapped or non-critical spin chains, the EE saturates to a constant value as the size of the subsystem increases, indicating short-range correlations and a finite correlation length. The slope of the EE concerning $\log(N)$ provides insights into the central charge, a key parameter characterizing the conformal field theory describing the quantum critical point.

Central charge values depend on various factors such as the quantum system's type, dimensionality, symmetry, and criticality. Although there is no universal formula for central charge values, some general features are observed for one-dimensional quantum systems, where central charge values are usually positive integers or fractions, reflecting the conformal symmetry of the system. 
Notably, different magnetic phases yield distinct central charge values in our study: $c = 1.44$ for SL-I, $c = 2.85$ for SL-II, and $c = 1.44$ for both phases in the presence of a magnetic field. These values differ from those of isotropic XX ($c = 1$) and anisotropic XY ($c = 0.5$) chain models \cite{EE-S-1,EE-S-2,EE-S-3}. The variation in central charge values in the SL-I and SL-II phases mirrors the different degrees of entanglement and conformal symmetry in the system. 
A higher central charge ($2.88$) in the SL-II phase aligns with the higher EE observed in Fig.~\ref{Fig3}. Larger central charge values suggest increased degrees of freedom and enhanced quantum correlations, emphasizing the TSI interaction's role in inducing novel quantum critical phenomena in the XX chain model. Furthermore, the critical SL-I and SL-II phases are distinctly signaled by their entanglement properties, particularly through different central charge values. The SL-I phase, surprisingly, deviates from the universality class typical for spin chains with next-nearest neighbor interactions \cite{EE-S-4} and one-dimensional spinless fermion models \cite{EE-S-5,EE-S-6,EE-S-7, EE-S-8, EE-S-9}, having a lower central charge of $c = 1.44$ instead of the expected $c = 1.5$. 
In contrast, the SL-II phase exhibits a higher central charge of $c = 2.88$, closely resembling $SU(N)$ spin models \cite{EE-S-10, EE-S-11}.

We have also studied the scaling behaviour of the SSP at the critical point ($\alpha_c=1.0$) between SL-I and SL-II phases, as shown in Fig.~\ref{Fig4} (c). We found that the SSP scales as $\xi s^2 \propto \sqrt{N}$, which is consistent with the SL-I phase.   This implies that the quantum coherence becomes weaker and more fragile as the system size increases. However, in the SL-II phase where $\alpha>\alpha_c=1$, the SSP shows a logarithmic scaling as $\xi s^2 \propto \ln (N)$.  This implies that the quantum entanglement becomes more robust and persistent as the system size increases. In the presence of a magnetic field, the data fluctuations prevent us from identifying any scaling behaviour.

\section{Conclusion}

In our research, we conducted a thorough investigation of the ground state properties in the spin-1/2 XX chain model. Specifically, we explored the impact of TSI and an external magnetic field. To gain insights into the system's quantum behaviors, we employed three fundamental metrics: coherency, spin squeezing, and entanglement entropy.
Our approach involved utilizing the fermionization technique to diagonalize the Hamiltonian and extract the exact ground state. Our findings revealed the presence of three distinct phases within the system: two gapless spin liquid phases, namely SL-I and SL-II, and a gapped PM phase. 
We utilized three essential tools namely, the $l_1$-norm of coherence, the SSP, and the EE to identify and characterize the quantum phase transitions between these phases.
In the absence of a magnetic field, the ground state exhibited non-squeezing, entanglement, and coherency in the SL-I and SL-II phases. Remarkably, the TSI showed minimal impact on these properties in the SL-I phase, signifying the existence of a distinctive gapless plateau state.
A critical quantum transition at the boundary separating SL-I and SL-II phases became evident through discontinuities in the first derivatives of these metrics.
The introduction of a magnetic field systematically reduced these properties and ultimately extinguished them within the PM phase.
Additionally, we identified a region characterized by fluctuations between squeezed and non-squeezed states. The border of this region, marked by coherent states indicated by SSP, left a discernible signature on the $l_1$-norm of coherence.
Our study provides valuable insights into the interplay of TSI and magnetic fields, shedding light on the intriguing quantum properties of the system.

We have also studied the scaling behavior of the $l_1$-norm of coherence, the SSP, and the EE with respect to the system size. We found that the $l_1$-norm of coherence behaves linearly in all states, in complete agreement with renormalization group predictions.

We obtained intriguing results regarding the EE. The EE can exhibit logarithmic divergence with respect to the system size at a quantum critical point, which is indicative of a departure from the area law. By examining the slope of the EE, we can determine the central charge, a crucial parameter that characterizes the conformal field theory governing the quantum critical point. What we found particularly interesting is that the EE deviated from the area law across the entire ground state phase diagram, indicating that both SL-I and SL-II phases exhibit critical behavior. In the absence of a magnetic field, the central charge in the SL-II phase was twice as large as in the SL-I phase. Additionally, in the presence of a magnetic field, the central charge exhibited the same behavior as in the SL-I phase.

In addition to investigating the EE, we explored the scaling behavior of the SSP. Intriguingly, we observed two distinct scaling behaviors in the absence of a magnetic field. In the critical SL-I phase, the SSP scales as $\propto \sqrt{N}$, while in the SL-II phase, it scales as $\propto \ln(N)$. 


\end{document}